\documentstyle[12pt,epsf]{ioplppt}
\begin{document}
\jl{1}
\eqnobysec
\def\qua{{
        \setlength{\unitlength}{0.1mm}
        \begin{picture}(24,20)(-2,1)
        \put(0,0){\line(1,0){20}}
        \put(0,0){\line(0,1){20}}
        \put(20,20){\line(-1,0){20}}
        \put(20,20){\line(0,-1){20}}
        \end{picture}
        }}
\def\tr{{\rm Tr\/}}
\def\FEF{Ginzburg-Landau potential}
\def\shock{bump}
\def\shocks{bumps}
%
%

\vspace*{0.0cm}
\thispagestyle{empty}
\begin{center}
   {\bf \Large 
Spontaneous Breaking of Translational Invariance and Spatial
Condensation in Stationary States on a Ring\\[2mm]
I. The Neutral System
 	}
   \\[13mm]
Peter F. Arndt$\mbox{}^{\star \S}$,
Thomas Heinzel$\mbox{}^\star$
and Vladimir Rittenberg$\mbox{}^{\star \dag}$
\\[7mm]
$\mbox{}^\star$
Physikalisches Institut \\Nu{\ss}allee 12,
53115 Bonn, Germany\\[3mm]
$\mbox{}^\dag$
SISSA\\ Via Beirut 2--4, 34014 Trieste, Italy\\[3mm]
$\mbox{}^\S$Institut f\"ur Theoretische Physik\\
Universit\"at zu K\"oln,
Z\"ulpicher Str. 77,
50937 K\"oln, Germany\\[1.0cm]
\end{center}
\renewcommand{\thefootnote}{\arabic{footnote}}
\addtocounter{footnote}{-1}
\vspace*{1mm}
%
We consider a model in which positive and negative particles with
equal densities diffuse in
an asymmetric, $CP$ invariant way on a ring. The positive particles hop
clockwise, the negative counter-clockwise and oppositely-charged adjacent
particles may swap positions. The model depends on two parameters.

Analytic calculations using quadratic algebras, inhomogeneous solutions of the
mean-field equations and Monte-Carlo simulations suggest that the model
has three phases. A pure phase in which one has three pinned blocks of
only positive, negative particles and vacancies and in which
translational invariance is broken. A mixed phase in which the current
has a linear dependence on one parameter but is independent of the
other one and of the density of the charged particles. In this phase one has
a bump and a fluid.
The bump (condensate)
contains positive and negative
particles only, the fluid contains charged
particles and vacancies uniformly distributed. 
The mixed phase is separated from the
disordered phase by a second-order phase-transition which has many properties of
 the
Bose-Einstein phase-transition observed in equilibrium.
Various critical exponents are found.
\vspace*{1.5cm}
\begin{flushleft}
\end{flushleft}
\thispagestyle{empty}
\mbox{}
%
%
%
%
\setcounter{page}{1}
\def\fI{
\begin{figure}[tb]
\setlength{\unitlength}{1mm}
\def\setl{\setlength\epsfxsize{10cm}}
\begin{picture}(80,70)
%
%
\put(20,0){
        \makebox{
                \setl
                \epsfbox{f1.epsf}}
        }
\put(20,65){\makebox{$j$}}
\put(100,16){\makebox{$\scriptscriptstyle \lambda=0.5$}}
\put(100,33){\makebox{$\scriptscriptstyle \lambda=1$}}
\put(100,53){\makebox{$\scriptscriptstyle \lambda=2$}}
\put(127,1){\makebox{$q$}}
\end{picture}
\caption{ 
The current in the large $L$ limit
as a function of $q$ for $\rho=0.2$, and
$\lambda=0.5,1,2$.
The curves are extrapolated from the data
of finite size lattices ($L\leq 200$) which are computed with the matrix product approach.
The solid line is given by $j=(q-1)/4$.
\label{fI}}
\end{figure}
}
\def\fIa{
\begin{figure}[tb]
\setlength{\unitlength}{1mm}
\def\setl{\setlength\epsfxsize{10cm}}
\begin{picture}(80,70)
%
%
\put(20,0){
        \makebox{
                \setl
                \epsfbox{f1a.epsf}}
        }
\put(20,65){\makebox{$\frac{4 J}{q-1}$}}
\put(100,23){\makebox{$\scriptscriptstyle \lambda=0.5$}}
\put(100,34){\makebox{$\scriptscriptstyle \lambda=1$}}
\put(100,50){\makebox{$\scriptscriptstyle \lambda=2$}}
\put(127,1){\makebox{$q$}}
\end{picture}
\caption{
The $q$ dependence of $4J/(q-1)$ for fixed $\rho=0.2$ and three values of
$\lambda$ ($\lambda=0.5,1$ and $2$). The data are obtained like in Fig. 1.
\label{fIa}}
\end{figure}
}
\def\fII{
\begin{figure}[tb]
\setlength{\unitlength}{1mm}
\def\setl{\setlength\epsfxsize{10cm}}
\begin{picture}(80,70)
%
\put(20,0){
        \makebox{
                \setl
                \epsfbox{f2.epsf}}
        }
\put(20,65){\makebox{$J$}}
\put(113,35){\makebox{$\scriptscriptstyle(a)$}}
\put(108,62){\makebox{$\scriptscriptstyle(b)$}}
\put(97,62){\makebox{$\scriptscriptstyle(c)$}}
\put(127,1){\makebox{$q$}}
\end{picture}
\caption{
The current $J$ as a function of $q$ for $\lambda=1$ and two densities
($\rho=0.2\; (+)$ and $\rho=0.4\; (\diamond)$)
The mean-field values are given by the dashed lines
$(a)$ ($\rho=0.2$) and $(b)$ ($\rho=0.4$).
The line $(c)$ is given by Eq.(\protect\ref{eqIII.V}).
The data are the results of extrapolations using the grand canonical ensemble and $L$ up to 200.
\label{fII}}
\end{figure}
}
\def\fIIa{
\begin{figure}[tb]
\setlength{\unitlength}{1mm}
\def\setl{\setlength\epsfxsize{10cm}}
\begin{picture}(80,70)
%
\put(20,0){
        \makebox{
                \setl
                \epsfbox{f2a.epsf}}
        }
\put(20,65){\makebox{$J_\pm, J$}}
\put(80,52){\makebox{$\scriptscriptstyle(a)$}}
\put(80,43){\makebox{$\scriptscriptstyle(c)$}}
\put(80,27){\makebox{$\scriptscriptstyle(b)$}}
\put(127,1){\makebox{$q$}}
\end{picture}
\caption{
The currents $J_+$ $(a)$, $J_-$ $(b)$ and the average current $J$
as a function of $q$ for $\lambda=1$, $p=0.4$ and $m=0.2$.
The points are obtained from Monte-Carlo simulations with $L=800$.
The solid lines $(a)$ and $(b)$ 
are given by Eq.(\protect\ref{eqIII.VII})
with $\eta=0.23$. The solid line $(c)$ is given by Eq.(\protect\ref{eqIII.V}).
\label{fIIa}}
\end{figure}
}
\def\fIIb{
\begin{figure}[tb]
\setlength{\unitlength}{1mm}
\def\setl{\setlength\epsfxsize{10cm}}
\begin{picture}(80,70)
%
\put(20,0){
        \makebox{
                \setl
                \epsfbox{f2b.epsf}}
        }
\put(20,65){\makebox{$j$}}
\put(77,45){\makebox{$\scriptscriptstyle(a)$}}
\put(77,26){\makebox{$\scriptscriptstyle(c)$}}
\put(77,11){\makebox{$\scriptscriptstyle(b)$}}
\put(127,1){\makebox{$q$}}
\end{picture}
\caption{
The current as a function of $q$ for $\lambda=1$ and asymmetric densities
$p=0.05$ and $m=0.35$. The points represent Monte-Carlo data for $L=400,800$ sites.
We show the current of negative particles (a), of positive particles (c)
and the average current (b).
The solid lines are given by the model (Eq. ):
$j=(1\pm\xi )(q-1)/4$ (a,c) with $\xi=0.6975$ (Eq. ) and
$j=(q-1)/4$ (b).
\label{fIIb}}
\end{figure}
}
\def\fIIc{
\begin{figure}[tb]
\setlength{\unitlength}{1mm}
\def\setl{\setlength\epsfxsize{10cm}}
\begin{picture}(80,70)
%
\put(20,0){
        \makebox{
                \setl
                \epsfbox{f2c.epsf}}
        }
\put(20,65){\makebox{$b$}}
\put(99,57){\makebox{$\scriptscriptstyle L=200$}}
\put(99,65){\makebox{$\scriptscriptstyle L=400$}}
\put(80,61){\makebox{\tiny model}}
\put(127,1){\makebox{$q$}}
\end{picture}
\caption{
The length of the fluid phase $b$ 
as a function of $q$ for $\lambda=1$ and asymmetric densities
$p=0.2$ and $m=0.4$. The points represent Monte-Carlo data for $L=200$ and $400$ sites.
The solid line is given by the model (Eq. ).
The phase transition is predicted for $q_c=29/17=1.70588$.
\label{fIIc}}
\end{figure}
}
\def\fIId{
\begin{figure}[tb]
\setlength{\unitlength}{1mm}
\def\setl{\setlength\epsfxsize{10cm}}
\begin{picture}(80,70)
%
\put(20,0){
        \makebox{
                \setl
                \epsfbox{f2d.epsf}}
        }
\put(20,65){\makebox{$v$}}
\put(99,63){\makebox{$\scriptscriptstyle L=400$}}
\put(99,70){\makebox{$\scriptscriptstyle L=800$}}
\put(127,1){\makebox{$q$}}
\end{picture}
\caption{
The length of the fluid phase $v$
as a function of $q$ for $\lambda=1$ and asymmetric densities
$p=0.2$ and $m=0.4$. The points represent Monte-Carlo data for $L=400$ and $800$ sites.
The solid line is given by the model (Eq. ).
The phase transition is predicted for $q_c=47/31=1.51613$.
\label{fIId}}
\end{figure}
}
\def\fIII{
\begin{figure}[tb]
\setlength{\unitlength}{1mm}
\def\setl{\setlength\epsfxsize{10cm}}
\begin{picture}(80,70)
%
\put(20,0){
        \makebox{
                \setl
                \epsfbox{f3.epsf}}
        }
\put(20,65){\makebox{$j$}}
\put(127,1){\makebox{$L$}}
\end{picture}
\caption{
The current as a function of $L$ for $q=0.8$, $\rho=0.2$, and $\lambda=1$. The data
are computed with the matrix product approach (bottom) and with Monte-Carlo simulations (top). There is an exponential approach 
to zero. Fitted are the curves $j=0.09 \exp(-0.055 L)$ and $j=0.25 \exp(-0.055 L)$.
\label{fIII}}
\end{figure}
}
\def\fIV{
\begin{figure}[tb]
\setlength{\unitlength}{1mm}
\def\setl{\setlength\epsfxsize{10cm}}
\begin{picture}(80,70)
%
\put(20,0){
        \makebox{
                \setl
                \epsfbox{f4.epsf}}
        }
\put(0,65){\makebox{$\scriptsize j(L)-j(L=\infty)$}}
\put(127,1){\makebox{$L$}}
\end{picture}
\caption{
The current as a function of $L$ for $q=1.2<q_c$, $\rho=0.2$, and $\lambda=1$. The data
are computed with the matrix product approach. We subtracted $j(L=\infty)=1/20$.
Fitted is the curve $j=1.6/L$. 
\label{fIV}}
\end{figure}
}
\def\fIVa{
\begin{figure}[tb]
\setlength{\unitlength}{1mm}
\def\setl{\setlength\epsfxsize{10cm}}
\begin{picture}(80,70)
%
\put(20,0){
        \makebox{
                \setl
                \epsfbox{f4a.epsf}}
        }
\put(8,65){\makebox{$\scriptsize \frac{J(L)-\frac{1}{20}}{\frac{1}{20}}$}}
\put(127,1){\makebox{$\scriptsize L$}}
\end{picture}
\caption{
The deviation of the current $J(L)$ for $L$ sites from its asymptotic
value 0.05 in the canonical ensemble ($q=1.2$, $\lambda=1$, $\rho=0.2$). The values obtained using Monte-Carlo
simulations ($+$) are shown together with those using the algebraic
approach ($\diamond$, up to $L=30$) and with those from inhomogeneous mean-field
solutions (dotted line) as described in \ref{appMF}.
\label{fIVa}}
\end{figure}
}
\def\fIVb{
\begin{figure}[tb]
\setlength{\unitlength}{1mm}
\def\setl{\setlength\epsfxsize{10cm}}
\begin{picture}(80,70)
%
\put(20,0){
        \makebox{
                \setl
                \epsfbox{f4b.epsf}}
        }
\put(3,65){\makebox{$\scriptsize \frac{J(L)-\frac{1}{20}}{\frac{1}{20}}$}}
\put(127,1){\makebox{$\scriptsize L$}}
\end{picture}
\caption{
The deviation of the current $J(L)$ for $L$ sites from its asymptotic
value 0.05 in the canonical ensemble ($q=1.2$, $\lambda=1$, $\rho=0.2$). The values obtained using Monte-Carlo
simulations ($+$) are shown together with those using the algebraic
approach ($\diamond$, up to $L=30$) and with those from inhomogeneous mean-field
solutions (line) as described in Appendix \ref{appMF}.

\label{fIVb}}
\end{figure}
}
\def\fV{
\begin{figure}[tb]
\setlength{\unitlength}{1mm}
\def\setl{\setlength\epsfxsize{10cm}}
\begin{picture}(80,70)
%
\put(20,0){
        \makebox{
                \setl
                \epsfbox{f5.epsf}}
        }
\put(-10,65){\makebox{$\scriptsize j(L)-j(L=\infty)$}}
\put(127,1){\makebox{$L$}}
\end{picture}
\caption{
The current as a function of $L$ for $q=2.2>q_c$, $\rho=0.2$, $\lambda=1$. The data
are computed with the matrix product approach. We subtracted $j(L=\infty)=0.18158$.
Fitted is the curve $j=0.04/L^2$.
\label{fV}}
\end{figure}
}
\def\fVI{
\begin{figure}[tb]
\setlength{\unitlength}{1mm}
\def\setl{\setlength\epsfxsize{10cm}}
\begin{picture}(80,70)
%
%
\put(20,0){
        \makebox{
                \setl
                \epsfbox{f6.epsf}}
        }
\put(20,65){\makebox{$j$}}
\put(127,1){\makebox{$L$}}
\end{picture}
\caption{
The current as a function of $L$ for the mean-field point
$\lambda=1$, $q=3$, $\rho=0.2$ and different ensembles.
One finds $j(L)=j(L=\infty)=1/5$ (grand canonical ensemble with at least one vacancy)
and $j(L)=1/5 L/(L-1)$ (canonical ensemble).
\label{fVI}}
\end{figure}
}
\def\fVII{
\begin{figure}[tb]
\setlength{\unitlength}{1mm}
\def\setl{\setlength\epsfxsize{10cm}}
\begin{picture}(80,70)
%
%
\put(20,0){
        \makebox{
                \setl
                \epsfbox{f7.epsf}}
        }
\put(-10,65){\makebox{$\scriptsize j(L=\infty)-j(L)$}}
\put(127,1){\makebox{$L$}}
\end{picture}
\caption{
The current as a function of $L$ for 
$q=3.8$, $\rho=0.2$, $\lambda=1$. The data
are computed with the matrix product approach.
One finds $j(L)=j(L=\infty)=0.21065$, and the approach is from below.
Fitted is the curve $j=0.016/L^2$.
\label{fVII}}
\end{figure}
}
\def\fVIII{
\begin{figure}[tb]
\setlength{\unitlength}{1mm}
\def\setl{\setlength\epsfxsize{10cm}}
\begin{picture}(80,70)
%
%
\put(20,0){
        \makebox{
                \setl
                \epsfbox{f8.epsf}}
        }
\put(20,65){\makebox{$c_{+,-}$}}
\put(127,1){\makebox{$y$}}
\end{picture}
\caption{
The correlation function $c_{+,-}$ obtained like $c_{0,0}$ in Fig.\ref{fVIII}.
\label{fVIII}}
\end{figure}
}
\def\fIX{
\begin{figure}[tb]
\setlength{\unitlength}{1mm}
\def\setl{\setlength\epsfxsize{10cm}}
\begin{picture}(80,70)
%
%
\put(20,0){
        \makebox{
                \setl
                \epsfbox{f9.epsf}}
        }
\put(20,65){\makebox{$c_{0,0}$}}
\put(127,1){\makebox{$y$}}
\end{picture}
\caption{
The correlation function $c_{0,0}$ obtained from
Monte-Carlo simulations
for $\rho=0.2$, $q=1.2$, $\lambda=1$, and
$L=400$. 
The solid line is obtained using Eq.(\ref{eqV.XXXIV}).
\label{fIX}}
\end{figure}
}
\def\fX{
\begin{figure}[tb]
\setlength{\unitlength}{1mm}
\def\setl{\setlength\epsfxsize{10cm}}
\begin{picture}(80,70)
%
%
\put(20,0){
        \makebox{
                \setl
                \epsfbox{f10.epsf}}
        }
\put(20,65){\makebox{$c_{+,+}$}}
\put(127,1){\makebox{$y$}}
\end{picture}
\caption{
The correlation $c_{+,+}$ for $\rho=0.2$, $q=1.2$, $\lambda=1$, and
$L=400$. The data are computed with Monte-Carlo simulations.
The solid line is the result from the phenomenological model (Eq.) where
$a=0.1175\pm 0.001$ and $b=8.129 \pm 0.02$ 
have been fitted from Monte-Carlo data from the condensates profiles.
\label{fX}}
\end{figure}
}
\def\fXa{
\begin{figure}[h]
\setlength{\unitlength}{1mm}
\def\setl{\setlength\epsfxsize{10cm}}
\begin{picture}(80,70)
%
%
\put(20,0){
        \makebox{
                \setl
                \epsfbox{f10a.epsf}}
        }
\put(22,64){\makebox{$c$}}
\put(125,-1){\makebox{$y$}}
\put(62,38){\makebox{$\scriptscriptstyle(a)$}}
\put(61,18){\makebox{$\scriptscriptstyle(b)$}}
\put(104,18){\makebox{$\scriptscriptstyle(c)$}}
\end{picture}
\caption{ \label{fXa}
The correlation functions $c_{0,0}$ $(a)$, $c_{+,-}$ $(b)$, and $c_{+,+}$ $(c)$
for $\lambda=1$, $q=0.5$, $\rho=0.2$, and
$L=100$.
The data are computed with Monte-Carlo simulations.
}
\end{figure}
}
\def\fXb{
\begin{figure}[tb]
\setlength{\unitlength}{1mm}
\def\setl{\setlength\epsfxsize{10cm}}
\begin{picture}(80,70)
%
%
\put(20,0){
        \makebox{
                \setl
                \epsfbox{f10b.epsf}}
        }
\put(22,64){\makebox{$c$}}
\put(125,-1){\makebox{$z$}}
\put(62,46){\makebox{$\scriptscriptstyle(a)$}}
\put(50,12){\makebox{$\scriptscriptstyle(b)$}}
\put(90,12){\makebox{$\scriptscriptstyle(c)$}}
\end{picture}
\caption{ \label{fXb}
The correlation functions $c_{0,0}$ $(a)$, $c_{+,-}$ $(b)$, and $c_{+,+}$ $(c)$
for $\lambda=1$, $q=2.2$, $\rho=0.2$, and
$L=100$.
The data are computed with Monte-Carlo simulations.
}
\end{figure}
}
\def\fXbb{
\begin{figure}[tb]
\setlength{\unitlength}{1mm}
\def\setl{\setlength\epsfxsize{10cm}}
\begin{picture}(80,70)
%
%
\put(20,0){
        \makebox{
                \setl
                \epsfbox{f10bb.epsf}}
        }
\put(20,65){\makebox{$\xi^{-1}$}}
\put(127,1){\makebox{$q-q_c$}}
\end{picture}
\caption{
The inverse correlation length as a function of $q$ for $\rho=0.2$, $\lambda=1$.
The data are obtained determining the correlation lengths from systems of
various lengths $L$ (up to 200) and then extrapolating the values to their
thermodynamical limit. The dashed curve is given by Eq.(\ref{eqVII.III}).
\label{fXbb}}
\end{figure}
}
\def\fXbbb{
\begin{figure}[h]
\setlength{\unitlength}{1mm}
\def\setl{\setlength\epsfxsize{10cm}}
\begin{picture}(80,70)
%
%
\put(20,0){
        \makebox{
                \setl
                \epsfbox{f10bbb.epsf}}
        }
\put(20,65){\makebox{$c_{+,-}$}}
\put(95,50){\makebox{$\scriptscriptstyle q=1.6$}}
\put(95,41){\makebox{$\scriptscriptstyle q=1.7$}}
\put(95,20){\makebox{$\scriptscriptstyle q=1.8$}}
\put(127,1){\makebox{$k$}}
\end{picture}
\caption{
The connected correlation function $c_{+,-}(k)$ as a function of $k$ for
for $\rho=0.2$, $\lambda=1$, $L=100$ and three values of $q$.
The data are computed using the grand canonical ensemble.
\label{fXbbb}}
\end{figure}
}
\def\fXc{
\begin{figure}[h]
\setlength{\unitlength}{1mm}
\def\setl{\setlength\epsfxsize{10cm}}
\begin{picture}(80,70)
%
%
\put(20,0){
        \makebox{
                \setl
                \epsfbox{f10c.epsf}}
        }
\put(22,64){\makebox{$c_{+,-}$}}
\put(125,-1){\makebox{$y$}}
\put(65,60){\makebox{$\scriptscriptstyle L=200$}}
\put(65,50){\makebox{$\scriptscriptstyle L=100$}}
\put(65,15){\makebox{$\scriptscriptstyle L=100$}}
\put(65,8){\makebox{$\scriptscriptstyle L=200$}}
\end{picture}
\caption{ \label{fXc}
The correlation functions $c_{+,-}$ for the canonical ensemble 
(bottom, computed by Monte-Carlo simulations) and for the grand canonical ensemble
(top) in the mixed phase:
$\lambda=1$, $q=1.2$, $\rho=0.2$,
$L=100$ and $L=200$ sites.
}
\end{figure}
}
\def\fXI{
\begin{figure}[tb]
\setlength{\unitlength}{1mm}
\def\setl{\setlength\epsfxsize{10cm}}
\begin{picture}(80,70)
%
%
\put(20,0){
        \makebox{
                \setl
                \epsfbox{f11.epsf}}
        }
\put(20,65){\makebox{$T$}}
\put(127,1){\makebox{$L$}}
\end{picture}
\caption{
The flip time for $\rho=0.2$, $q=0.8$, $\lambda=1$
(exponential increase in the pure phase).
The data are computed with Monte-Carlo simulations.
The solid line is the time the local density of
vacancies stays at the low peak of the density distribution,
the broken line is the time the local density stays at the high peak
of the distribution.
\label{fXI}}
\end{figure}
}
\def\fXII{
\begin{figure}[tb]
\setlength{\unitlength}{1mm}
\def\setl{\setlength\epsfxsize{10cm}}
\begin{picture}(80,70)
%
%
\put(20,0){
        \makebox{
                \setl
                \epsfbox{f12.epsf}}
        }
\put(20,65){\makebox{$T$}}
\put(127,1){\makebox{$L$}}
\end{picture}
\caption{
The flip time for $\rho=0.2$, $q=1.2$, $\lambda=1$
(algebraic increase).
The data are computed with Monte-Carlo simulations.
The solid line is the time the local density of
vacancies stays at the low peak of its probability distribution,
the broken line is the time the local density stays at the high peak
of its distribution.
\label{fXII}}
\end{figure}
}
\def\fXIII{
\begin{figure}[tb]
\setlength{\unitlength}{1mm}
\def\setl{\setlength\epsfxsize{10cm}}
\begin{picture}(80,70)
%
%
\put(20,0){
        \makebox{
                \setl
                \epsfbox{f13.epsf}}
        }
\put(18,67){\makebox{$c(1)$}}
\put(48,10){\makebox{$\scriptstyle c_{+,0}$}}
\put(41,15){\makebox{$\scriptstyle c_{0,+}$}}
\put(45,39){\makebox{$\scriptstyle c_{-,+}$}}
\put(53,41){\makebox{$\scriptstyle c_{+,+}$}}
\put(43,56){\makebox{$\scriptstyle c_{+,-}$}}
\put(53,62){\makebox{$\scriptstyle c_{0,0}$}}
\put(127,1){\makebox{$q$}}
\end{picture}
\caption{
Large $L$ limit of the connected nearest neighbour 
correlations $c_{+,0}(1)=c_{0,-}(1)$, $c_{0,+}(1)=c_{-,0}(1)$,
$c_{-,+}(1)$, $c_{+,+}(1)=c_{-,-}(1)$, $c_{+,-}(1)$ and $c_{0,0}(1)$
for $\lambda=1$ and $\rho=0.2$.
The data are the results of extrapolations using the grand canonical ensemble and $L$ up to 150.
\label{fXIII}}
\end{figure}
}
\def\fXIV{
\begin{figure}[tb]
\setlength{\unitlength}{1mm}
\def\setl{\setlength\epsfxsize{10cm}}
\begin{picture}(80,70)
%
%
\put(20,0){
        \makebox{
                \setl
                \epsfbox{f14.epsf}}
        }
\put(20,65){\makebox{$T$}}
\put(127,1){\makebox{$L$}}
\end{picture}
\caption{
The average flip time $T$ defined in the text as a function of L in the
pure phase ($\rho=0.2$, $\lambda=1$, $q=0.8$) obtained from Monte-Carlo
simulations.
\label{fXIV}}
\end{figure}
}
\def\fXV{
\begin{figure}[tb]
\setlength{\unitlength}{1mm}
\def\setl{\setlength\epsfxsize{10cm}}
\begin{picture}(80,70)
%
%
\put(20,0){
        \makebox{
                \setl
                \epsfbox{f15.epsf}}
        }
\put(20,65){\makebox{$T$}}
\put(127,1){\makebox{$L$}}
\end{picture}
\caption{
Same as in Fig.\ref{fXIV} for the mixed phase ($q=1.2$).
\label{fXV}}
\end{figure}
}
\def\fXVI{
\begin{figure}[tb]
\setlength{\unitlength}{1mm}
\def\setl{\setlength\epsfxsize{10cm}}
\begin{picture}(80,70)
%
%
\put(20,0){
        \makebox{
                \setl
                \epsfbox{f16.epsf}}
        }
\put(23,65){\makebox{$\zeta$}}
\put(127,1){\makebox{$L$}}
\end{picture}
\caption{
$\zeta$ defined by Eq.(\ref{eqVIII.II}) as a function of $L$ for $\rho=0.2$, $\lambda=1$ for $q=0.9$,
$1.4$, $1.6$ and $2.2$ (from top to bottom).
The data are obtained using the grand canonical ensemble.
\label{fXVI}}
\end{figure}
}
\def\fXVII{
\begin{figure}[tb]
\setlength{\unitlength}{1mm}
\def\setl{\setlength\epsfxsize{10cm}}
\begin{picture}(80,70)
%
%
\put(20,0){
        \makebox{
                \setl
                \epsfbox{f17.epsf}}
        }
\put(15,65){\makebox{$f(|M|)$}}
\put(127,1){\makebox{$|M|$}}
\end{picture}
\caption{
The \FEF\ in the mixed phase ($\rho=0.2$,
$\lambda=1$, $q=1.2$) as a function of $|M|$ obtained from Monte-Carlo simulations
for $L=200,\, 400,\, 600,\, 800,\, 1000$ (from top to bottom).
\label{fXVII}}
\end{figure}
}
\def\fXVIII{
\begin{figure}[h]
\setlength{\unitlength}{1mm}
\def\setl{\setlength\epsfxsize{10cm}}
\begin{picture}(80,70)
%
%
\put(20,0){
        \makebox{
                \setl
                \epsfbox{f18new.epsf}}
        }
\put(12,65){\makebox{$\frac1{L}f(|M|)$}}
\put(127,1){\makebox{$|M|$}}
\end{picture}
\caption{
The \FEF\ in the pure phase ($\rho=0.2$,
$\lambda=1$, $q=0.9$) as a function of $|M|$ obtained from Monte-Carlo simulations
for $L=100$($\diamond$), $200$(+) and $400$($\protect\qua$).
\label{fXVIII}}
\end{figure}
}
\def\fXIX{
\begin{figure}[tb]
\setlength{\unitlength}{1mm}
\def\setl{\setlength\epsfxsize{10cm}}
\begin{picture}(80,70)
%
%
\put(20,0){
        \makebox{
                \setl
                \epsfbox{f19.epsf}}
        }
\put(15,65){\makebox{$f(|M|)$}}
\put(127,1){\makebox{$|M|$}}
\end{picture}
\caption{
The \FEF\ in the disordered phase ($\rho=0.2$,
$\lambda=1$, $q=2.5$) as a function of $|M|$ obtained from Monte-Carlo simulations
for $L=400,\, 600,\, 800 ,\, 1000,\, 1200,\, 1400$ (from top to bottom).
\label{fXIX}}
\end{figure}
}
\def\fXX{
\begin{figure}[h]
\setlength{\unitlength}{1mm}
\def\setl{\setlength\epsfxsize{10cm}}
\begin{picture}(80,70)
%
%
\put(20,0){
        \makebox{
                \setl
                \epsfbox{f20.epsf}}
        }
\put(20,65){\makebox{$\lambda$}}
\put(48,50){\makebox{$\scriptscriptstyle(a)$}}
\put(38,45){\makebox{\tiny pure phase}}
\put(57,45){\makebox{\tiny mixed phase}}
\put(77,20){\makebox{\tiny disordered phase}}
\put(84,50){\makebox{$\scriptscriptstyle(b)$}}
\put(103,28){\makebox{$\scriptscriptstyle(c)$}}
\put(103,21){\makebox{$\scriptscriptstyle(d)$}}
\put(103,15){\makebox{$\scriptscriptstyle(e)$}}
\put(103,10){\makebox{$\scriptscriptstyle(f)$}}
\put(127,1){\makebox{$q$}}
\end{picture}
\caption{
Phase diagram in the $q$-$\lambda$-plane for $\rho=0.2$: 
The points indicate measured transition points between
pure/mixed and mixed/disordered phase (compare table \protect\ref{tI}). 
Also shown
are the lines $(a)$: $q=1$ (expected transition pure/mixed phase) and
$(b)$: $q=(0.8\lambda+1.4)/1.4$ (transition mixed/disordered phase as predicted
by the model Eq. ). 
In the disordered phase we also indicate where 
representations of the algebra exist (Eq.) with dimensions 1 (line $(c)$\/), 
2 (line$(d)$\/), 3 (line $(e)$\/), 4 (line$(f)$\/).
\label{fXX}}
\end{figure}
}
\def\fXXI{
\begin{figure}[h]
\setlength{\unitlength}{1mm}
\def\setl{\setlength\epsfxsize{10cm}}
\begin{picture}(80,70)
%
%
\put(20,0){
        \makebox{
                \setl
                \epsfbox{f21.epsf}}
        }
\put(20,65){\makebox{$\rho_{\rm fluid}$}}
\put(52,19){\makebox{$\scriptscriptstyle \rho_c$}}
\put(100,62){\makebox{\tiny model}}
\put(104,44){\makebox{$\scriptscriptstyle \rho=0.2$}}
\put(107,53){\makebox{$\scriptscriptstyle \rho=0.4$}}
\put(127,1){\makebox{$q$}}
\end{picture}
\caption{Independence of the
density in the fluid phase on the system's overall density:
The data is computed with Monte-Carlo simulations for $\rho=0.2$ and $\rho=0.4$
for $L=400$ sites. We also show the critical density $\rho_c$ computed with the grand
canonical ensemble and the curve of the density of the fluid phase
as predicted by the model (Eq. ). 
\label{fXXI}}
\end{figure}
}
\def\fXXII{
\begin{figure}[tb]
\setlength{\unitlength}{1mm}
\def\setl{\setlength\epsfxsize{10cm}}
\begin{picture}(80,70)
%
%
\put(20,0){
        \makebox{
                \setl
                \epsfbox{f22.epsf}}
        }
\put(20,65){\makebox{$\rho$}}
\put(66,30){\makebox{$\scriptscriptstyle L=200$}}
\put(78,30){\makebox{$\scriptscriptstyle 150$}}
\put(84,30){\makebox{$\scriptscriptstyle 100$}}
\put(127,1){\makebox{$z$}}
\end{picture}
\caption{
The density $\rho$ as a function of the fugacity $z$ for $q=1.2$, $\lambda=1$, $L=100,150,200$ sites.
\label{fXXII}}
\end{figure}
}
\def\fXXIII{
\begin{figure}[tb]
\setlength{\unitlength}{1mm}
\def\setl{\setlength\epsfxsize{10cm}}
\begin{picture}(80,70)
%
%
\put(20,0){
        \makebox{
                \setl
                \epsfbox{f23.epsf}}
        }
\put(20,65){\makebox{$a$}}
\put(66,30){\makebox{$\scriptscriptstyle L=200$}}
\put(66,9){\makebox{$\scriptscriptstyle L=400$}}
\put(127,1){\makebox{$q$}}
\end{picture}
\caption{
The parameter $a$ of the solution $a \tan(b z)$ of the phenomenological equation (Eq.) for
the Bose condensate as fitted from Monte-Carlo data
for $L=200,400$.
\label{fXXIII}}
\end{figure}
}
\def\fXXIV{
\begin{figure}[tb]
\setlength{\unitlength}{1mm}
\def\setl{\setlength\epsfxsize{10cm}}
\begin{picture}(80,70)
%
%
\put(20,0){
        \makebox{
                \setl
                \epsfbox{f24.epsf}}
        }
\put(20,65){\makebox{$b$}}
\put(66,28){\makebox{$\scriptscriptstyle L=200$}}
\put(66,18){\makebox{$\scriptscriptstyle L=400$}}
\put(127,1){\makebox{$q$}}
\end{picture}
\caption{
The parameter $b$ of the solution $a \tan(b z)$ of the phenomenological equation (Eq.) for
the Bose condensate as fitted from Monte-Carlo data
for $L=200,400$.
\label{fXXIV}}
\end{figure}
}
\def\fXXV{
\begin{figure}[h]
\setlength{\unitlength}{1mm}
\def\setl{\setlength\epsfxsize{10cm}}
\begin{picture}(80,70)
%
%
\put(20,0){
        \makebox{
                \setl
                \epsfbox{f25.epsf}}
        }
\put(20,65){\makebox{$c_+,c_-$}}
\put(127,1){\makebox{$y$}}
\end{picture}
\caption{
Density profiles in the Bose-Einstein-condensate for $q=1.2$ and $\lambda=1$: \protect\\
a. $a \tan(b z)$ from the phenomenological model where $a=$ and $b=$ have been fitted from the Monte-Carlo data\protect\\
b. inhomogeneous mean-field solution ($L=200$) \protect\\
c. Monte-Carlo data ($L=200$).
\label{fXXV}}
\end{figure}
}
\def\fXXVI{
\begin{figure}[tb]
\setlength{\unitlength}{1mm}
\def\setl{\setlength\epsfxsize{10cm}}
\begin{picture}(80,70)
%
%
\put(20,0){
        \makebox{
                \setl
                \epsfbox{f26.epsf}}
        }
\put(20,65){\makebox{$c_+,c_-$}}
\put(127,1){\makebox{$y$}}
\end{picture}
\caption{
The density profile in the Bose-condensate for $q=1$ and $\lambda=1$: \protect\\
points: Monte-Carlo data ($L=200$).\protect\\
lines: inhomogeneous mean-field solution ($L=200$)
\label{fXXVI}}
\end{figure}
}
\def\fXXVII{
\begin{figure}[tb]
\setlength{\unitlength}{1mm}
\def\setl{\setlength\epsfxsize{10cm}}
\begin{picture}(80,70)
%
%
\put(20,0){
        \makebox{
                \setl
                \epsfbox{f27.epsf}}
        }
\put(20,65){\makebox{$\xi$}}
\put(127,1){\makebox{$q$}}
\end{picture}
\caption{
The correlation length $\xi$ of the pure phase
as measured from the decay of the correlation function $c_{+,-}(k)=A \exp(-k/\xi)+p\, m$.
The data is computed with the matrix product approach for $\rho=0.2$ and $L$ up to 200,
followed by extrapolation to estimate the $L\rightarrow\infty$ limit.
\label{fXXVII}}
\end{figure}
}
\def\fXXVIII{
\begin{figure}[tb]
\setlength{\unitlength}{1mm}
\def\setl{\setlength\epsfxsize{10cm}}
\begin{picture}(80,70)
%
%
\put(20,0){
        \makebox{
                \setl
                \epsfbox{f28.epsf}}
        }
\put(22,64){\makebox{$c_{+,-}$}}
\put(125,-1){\makebox{$y$}}
\end{picture}
\caption{
The correlation function $c_{+,-}$
as a function of $y$ for $\rho=0.2$, $\lambda=1$
$q=1.2$, $L=100$ and various values of $\nu=0,0.05,0.06,0.07,0.08$ (from top to bottom).
The solid line corresponds to the fluid (see the text).
\label{fXXVIII}}
\end{figure}
}
\def\fXXIX{
\begin{figure}[tb]
\setlength{\unitlength}{1mm}
\def\setl{\setlength\epsfxsize{10cm}}
\begin{picture}(80,70)
%
%
%
\put(20,0){
        \makebox{
                \setl
                \epsfbox{f29.epsf}}
        }
\put(23,64){\makebox{$\zeta$}}
\put(125,-1){\makebox{$L$}}
\end{picture}
\caption{
$\zeta$ defined by Eq.(\ref{eqVIII.II}) as a function of $L$ for $\rho=0.2$, $\lambda=1$ and
$q=1.2$ for various values of $\nu$ (from top to bottom $\nu=0,0.05,0.06,0.07,0.08,0.09,0.1$)
\label{fXXIX}}
\end{figure}
}
\def\fXXX{
\begin{figure}[h]
\setlength{\unitlength}{1mm}
\def\setl{\setlength\epsfxsize{10cm}}
\begin{picture}(80,70)
%
%
\put(20,0){
        \makebox{
                \setl
                \epsfbox{f30.epsf}}
        }
\put(22,64){\makebox{$P$}}
\put(125,-1){\makebox{$1/\rho$}}
\end{picture}
\caption{
The ``pressure'' $P$ defined by Eq.(\protect\ref{eqVIII.III}) as a function of $1/\rho$ for $q=1.2$ and
$\lambda=1$. The data are obtained as explained in the text.
\label{fXXX}}
\end{figure}
}
\def\fXXXI{
\begin{figure}[tb]
\setlength{\unitlength}{1mm}
\def\setl{\setlength\epsfxsize{10cm}}
\begin{picture}(80,70)
%
%
\put(20,0){
        \makebox{
                \setl
                \epsfbox{f31.epsf}}
        }
\put(22,64){\makebox{$c$}}
\put(125,-1){\makebox{$y$}}
\put(62,38){\makebox{$\scriptscriptstyle (a)$}}
\put(62,18){\makebox{$\scriptscriptstyle (b)$}}
\end{picture}
\caption{ \label{fXXXI}
The $y$ dependence of the two-point correlation functions $c_{0,0}$  $(a)$, 
$c_{+,-}$ $(b)$ 
in the pure phase ($q=0.5$, $\lambda=1$, $\rho=0.2$) 
and lattice sizes $L=75 (\diamond,\protect\qua)$ 
and $L=100 (+,\times)$.
(Monte-Carlo simulations).
}
\end{figure}
}
\def\fInew{
\begin{figure}[tb]
\setlength{\unitlength}{1mm}
\def\setl{\setlength\epsfxsize{10cm}}
\begin{picture}(80,70)
%
%
%
\put(20,0){
        \makebox{
                \setl
                \epsfbox{f1new.epsf}}
        }
\put(40,50){\makebox{$\rho_+^{\rm co}$}}
\put(108,50){\makebox{$\rho_-^{\rm co}$}}
\put(127,1){\makebox{$y$}}
\end{picture}
\caption{
The concentrations of positive and negative particles in a shock ($q=1.2$,
$\lambda=1$, $\rho=0.2$, $L=400$) obtained from Monte-Carlo simulations and from the model
as described in the text (solid curves).
\label{fInew}}
\end{figure}
}
\def\fIInew{
\begin{figure}[tb]
\setlength{\unitlength}{1mm}
\def\setl{\setlength\epsfxsize{10cm}}
\begin{picture}(80,70)
%
%
%
\put(20,0){
        \makebox{
                \setl
                \epsfbox{f2new.epsf}}
        }
\put(50,50){\makebox{$\rho_+^{\rm co}$}}
\put(100,50){\makebox{$\rho_-^{\rm co}$}}
\put(127,1){\makebox{$y$}}
\end{picture}
\caption{
Like in Fig.\ref{fInew} except that here $q=1$.
\label{fIInew}}
\end{figure}
}
\def\fC{
\begin{figure}[h]
\setlength{\unitlength}{1mm}
\def\setl{\setlength\epsfxsize{10cm}}
\begin{picture}(80,82)
\put(20,0){
        \makebox{
                \setl
                \epsfbox{figC.epsf}}
        }
\put(20,65){\makebox{$J$}}
\put(127,1){\makebox{$\rho$}}
\end{picture}
\caption{ \label{fC}
The current as a function of the density. 
The data are obtained from Monte-Carlo simulations for $q=1.2$ and $\lambda=1$ and
$L=100$ (top), $L=200$ and $L=400$ (bottom).
The horizontal line gives the value of the current in the mixed phase (see Eq.(\ref{eqIII.V})).
}
\end{figure}
}
\def\f78{
\begin{figure}[h]
\setlength{\unitlength}{1mm}
\def\setl{\setlength\epsfxsize{10cm}}
\begin{picture}(80,82)
\put(20,0){
        \makebox{
                \setl
                \epsfbox{fig78.epsf}}
        }
\put(55,46){\makebox{$\scriptscriptstyle(a)$}}
\put(50,14){\makebox{$\scriptscriptstyle(b)$}}
\put(22,64){\makebox{$c$}}
\put(127,1){\makebox{$y$}}
\end{picture}
\caption{ \label{f78}
The correlation functions $c_{0,0}$ $(a)$ and $c_{+,-}$ $(b)$ obtained from
Monte-Carlo simulations
for $\rho=0.2$, $q=1.2$, $\lambda=1$, and
$L=400$. 
The solid lines are obtained using the mean-field results presented in Sec.\ref{secIMF}. 
}
\end{figure}
}
\def\fCI{
\begin{figure}[h]
\setlength{\unitlength}{1mm}
\def\setl{\setlength\epsfxsize{10cm}}
\begin{picture}(80,82)
\put(20,0){
        \makebox{
                \setl
                \epsfbox{f101.epsf}}
        }
\put(35,39){\makebox{$\scriptscriptstyle(a)$}}
\put(35,23){\makebox{$\scriptscriptstyle(b)$}}
\put(35,11){\makebox{$\scriptscriptstyle(c)$}}
\put(43,46){\makebox{$\scriptscriptstyle(d)$}}
\put(45,28){\makebox{$\scriptscriptstyle(e)$}}
\put(55,16){\makebox{$\scriptscriptstyle(f)$}}
\put(7,64){\makebox{$f(v_l)$}}
\put(127,1){\makebox{$v_l$}}
\end{picture}
\caption{ \label{fCI}
The \FEF\ $f(v_l)$ as a function of the local density of vacancies $v_l$.
The curves are for $q=1.6\approx q_c$ and $L=400 (a),L=800 (b), L=1600 (c)$, and
for $q=2.2$ and $L=400 (d),L=800 (e), L=1600 (f)$. In all cases we have taken $\lambda=1$ and $\rho=0.2$.
}
\end{figure}
}
\def\fCX{
\begin{figure}[h]
\setlength{\unitlength}{1mm}
\def\setl{\setlength\epsfxsize{10cm}}
\begin{picture}(80,82)
\put(20,0){
        \makebox{
                \setl
                \epsfbox{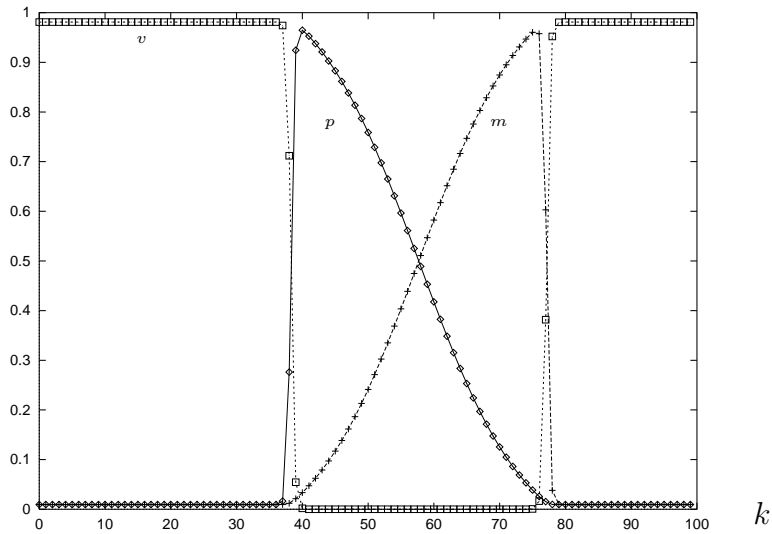}}
        }
\put(45,65){\makebox{$\scriptscriptstyle v $}}
\put(70,54){\makebox{$\scriptscriptstyle p $}}
\put(92,54){\makebox{$\scriptscriptstyle m $}}
\put(22,64){\makebox{$$}}
\put(127,1){\makebox{$k$}}
\end{picture}
\caption{ \label{fCX}
The inhomogeneous solution of the mean-field equations in the pure phase for $q=0.9,\, \lambda=1,\, p=m=0.2$ and $L=100$.
}
\end{figure}
}
\def\fCXI{
\begin{figure}[h]
\setlength{\unitlength}{1mm}
\def\setl{\setlength\epsfxsize{10cm}}
\begin{picture}(80,82)
\put(20,0){
        \makebox{
                \setl
                \epsfbox{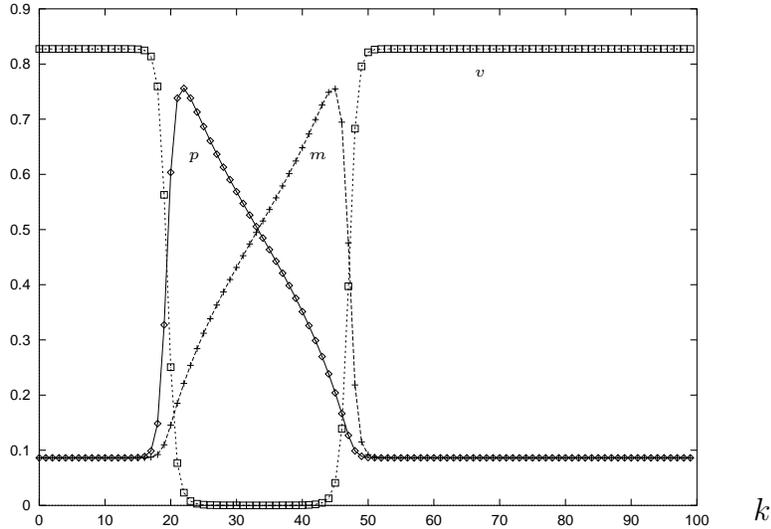}}
        }
\put(90,60){\makebox{$\scriptscriptstyle v $}}
\put(52,49){\makebox{$\scriptscriptstyle p $}}
\put(68,49){\makebox{$\scriptscriptstyle m $}}
\put(22,64){\makebox{$$}}
\put(127,1){\makebox{$k$}}
\end{picture}
\caption{ \label{fCXI}
Same as in Fig.\ref{fCX} for the mixed phase for $q=1.2,\, \lambda=1,\, p=m=0.2$ and $L=100$.
}
\end{figure}
}
\def\tI{
\begin{table}[tb]
\caption{
The estimates of the critical value of $q$ obtained using the grand canonical ensemble ($q_c^{gc}$)
or Monte-Carlo simulations ($q_c^{c}$) compared
with the values $q_c$ given by formula (\protect\ref{eqV.XXIX}).
\protect\label{tI}
\protect\label{tabI}
}
\begin{indented}
\item[]
\begin{tabular}{@{}lllllllllll}
\br
&\centre{2}{$p=m=0.1$}&&\centre{2}{$p=m=0.2$}&&\centre{2}{$p=m=0.4$} \\
&\crule{2}&&\crule{2}&&\crule{2}\\
$\lambda$\;\;&$q_c^{gc}$&$q_c$&\;&$q_c^{gc}$&$q_c$&\;&$q_c^{gc}$&$q_c$\\
\mr
0.25& 1.10&1.08		&&1.2 &1.14	&&1.3 &1.22&\\
0.50& 1.15&1.17		&&1.3 &1.29	&&1.6 &1.44&\\
0.75& 1.25 &1.25	&&1.5 &1.42  	&&1.8&1.67&\\
1.00& 1.35 &1.33	&&1.62&1.57	&&2.2 &1.89&\\
2.00& 1.60 &1.67	&&2.2 &2.14	&&2.9 &2.78&\\
3.00& 1.85 &2.00  	&&2.5 &2.71	&&3.5 &3.67&\\
4.00& 2.0  &2.33  	&&2.8 &3.28	&&3.8 &4.56&\\
\br
\end{tabular}
\end{indented}
\end{table}
}
\section{Introduction\label{secI}}

While the conceptual framework for the understanding of spontaneous
symmetry breaking (SSB) occurring in systems at thermal equilibrium is well
established, it is much poorer for systems far from equilibrium. It was
realized recently though, that exclusion models with hard core 
interaction, are simple enough to allow analytic methods, either
approximative or exact, for the description of their stationary states,
yet exhibit a rich variety of behaviors even in one dimension. For
example SSB of $CP$ symmetry in a first-order phase transition was observed
in open systems with two \cite{citR,citS,citO} or three species \cite{citZ,citU}.

  The aim of the present work is to analyze the structure of the
stationary states of a one-dimensional three-species exclusion model on a
ring with continuous symmetries and, in particular, to address the question
of the existence of SSB. In this model, introduced in \cite{citB}, positive and
negative particles diffuse in an asymmetric, $CP$ invariant way. The positive
particles hop clockwise, the negative particles counter-clockwise and
oppositely charged adjacent particles may exchange positions. The model
thus defined is translationally invariant and the numbers of positive and
negative particles are conserved. The study presented in \cite{citB} suggests the
existence of three phases in the stationary state. 
In particular, as we shall see, for two of these
phases  charge segregation (segregation of species) occurs in the system.
Here we confirm these predictions, we provide a more complete analysis of
these phases and try to unravel the mechanisms by which phase transitions
occur in such a simple system. A summary of the features of the phases
and of the phenomenology of the model is given below. We first come back
to its definition. 

  We consider a ring of $L$ sites, labeled by $k=1,\dots,L$. Each site is
occupied by one of the three types of particles, 0, 1 or 2. The particle of
type 0 is hereafter referred to as a vacancy. The dynamics of the model is
defined as follows. At each time step a bond is chosen randomly, and the
particles $\alpha$ and $\beta$ located at the two ends of the bond exchange
positions with the rate $g_{\alpha,\beta}$, where the indices $\alpha,\,\beta=0,1,2$.
Therefore, the dynamics conserves the number of particles $N_\alpha$ of each species.
Up to this point, this defines a class of models depending on five
parameters, since rates are defined up to a rescaling of time. Here we
restrict the model taking the following rates
\begin{eqnarray}
(1)(0) \rightarrow (0)(1)
\quad\quad
&\mbox{with rate }g_{1,0}=\lambda
\nonumber\\
(0)(2) \rightarrow (2)(0)
&\mbox{with rate }g_{0,2}=\lambda
\nonumber\\
(2)(1) \rightarrow (1)(2)
&\mbox{with rate }g_{2,1}=1      
\nonumber\\
(1)(2) \rightarrow (2)(1)
&\mbox{with rate }g_{1,2}=q      
\label{eqI.I}
\end{eqnarray}
i.e. the remaining rates $g_{0,1}$ and $g_{2,0}$ are zero. The model depends on
two parameters $\lambda$ and $q$ (for convenience, we introduce also the
parameter $r=1/q$). In order to facilitate the intuitive description of the
model, we will also denote the particles of type 1 and 2 by $(+)$ and $(-)$
respectively and rewrite (\ref{eqI.I}) as
\begin{eqnarray}
g_{+,0}=g_{0,-}=\lambda,
\qquad
g_{-,+}=1,
\qquad
g_{+,-}=q
\label{eqI.II}
\end{eqnarray}
In what follows we will consider only the case of
equal densities of charged particles that we are going to denote by $\rho$.
With the choice of rates (\ref{eqI.II}), we expect therefore the stationary state to
be $CP$ invariant and the currents of positive and negative particles to be
equal.
The case of unequal densities is subject of a separate paper \cite{citQ}.

  For the sake of clarity, we present here a summary of the main
phenomenological features of this model, using a simple intuitive
presentation. Our knowledge is based on the study of the model by exact
algebraic methods, mean-field analysis and Monte-Carlo simulations,
described in the later sections. For a given density $\rho$, the phase
diagram of the model consists of three sectors in the $q-\lambda$ plane.

The {\em pure phase}. ($0<q<1$, any $\lambda$ and $\rho$). If one takes a
system of size $L$ and start from an arbitrary configuration, in the long
time regime, the system organizes itself into three blocks
\begin{eqnarray}
(\,0\,) \cdots (\,0\,) (+) \cdots (+) (-) \cdots (-)
\label{eqI.III}
\end{eqnarray}
building what we are going to call a \shock\ which covers the whole ring.
This may be simply understood looking at the rates (\ref{eqI.II}). Since $q<1$, in
the average, particles of type $(-)$ will be preferentially located to the
right of particles of type $(+)$. Themselves will be found to the right of
particles of type $(\,0\,)$, while the latter will be found on the ring to the
right of particles of type $(-)$. The average waiting time between hops of
the \shocks\ from one position on the ring to another one increase
exponentially with $L$. Let us give an intuitive explanation of this point.
Prepare the system as in (\ref{eqI.III}). According to (\ref{eqI.II}) the single bond that is
allowed to evolve is the one located at the interface between the positive
and negative blocks. Consider the positive particle at the interface. It
will begin diffusing through the negative block to its right, but with a
bias to the left since $q<1$, resulting in a restoring force on the
particle. As a consequence, the probability for this particle to
transverse the negative block is exponentially decreasing with the size of
the system. In the thermodynamic limit, any configuration (\ref{eqI.III}) on the
ring is an absorbing state of the dynamics, i.e. the system does not
evolve. Otherwise stated, the configuration space is decomposed into a
continuous infinity of ``ergodic'' components, i.e. ergodicity is infinitely
broken. As a consequence, in the thermodynamic limit, translational
invariance is broken, the current is zero and the segregation of charges
maximal, corresponding to a complete charge segregation (hence the name
pure phase). Let us observe that, in general, if one denotes by
$c_{\alpha,\beta}(R,L)$ any connected two-point function for a system of size $L$
and if instead of the distance $R$ one takes the dimensionless variable
$y=R/L$, charge segregation implies that the limits
\begin{eqnarray}
\lim_{L\rightarrow\infty} c_{\alpha,\beta}(R,L) =
\lim_{L\rightarrow\infty} c_{\alpha,\beta}(y,L) =
c_{\alpha,\beta}(y)
\label{eqI.IV}
\end{eqnarray}
exist (similar limits exist also for any $n$-point functions). Charge
segregation was seen in two \cite{citR,citO,citS,citZA} and three-species \cite{citU} open systems
when first-order phase transitions take place. In the present case the
functions $c_{\alpha,\beta}(y)$ have a simple expression which can be computed
using (\ref{eqI.III}).
The pure phase was seen also in Ref.\cite{citA} for $\rho=1/3$ in a different model
where the rates are
\begin{eqnarray}
g_{1,0}=
g_{0,2}=
g_{2,1}=1,\qquad
g_{0,1}=
g_{2,0}=
g_{1,2}
\label{eqI.V}
\end{eqnarray}
The finite-size behavior of the present model and that of Ref.\cite{citA} are
different.

The {\em mixed phase}. For fixed $\lambda$ and $\rho$, this phase appears
as soon as $1<q$ and survives until $q$ reaches a critical value 
\begin{eqnarray}
q_c=1+\frac{4 \lambda \rho}{1+2\rho}
\label{eqI.VI}
\end{eqnarray}
Again segregation of species is observed, though differently from what was
observed previously for $q<1$. In order clarify the physics let us take $q$
larger but close to~1. The two blocks of $(+)$ and $(-)$ particles, which
were distinct in the pure phase, merge in one region, hereafter called
condensate. The rest
of the system, hereafter called fluid, is mainly occupied by vacancies and
traversed by a few $(+)$ and $(-)$ particles distributed in an uniform way.
This may be understood as follows. Prepare the system in the configuration
(\ref{eqI.III}). The only particles that can move are located at the $(+)(-)$ interface.
The difference is that now the bias, proportional to $q-1$, is positive,
i.e. oriented to the right. Therefore the $(+)$ particle has a finite
probability to traverse the negative block on its right. As soon as a $(+)$
particle leaves the condensate at its right end, it reappears almost
instantly at its left end. In other words the positive component of
the condensate melts on the right but is restored on the left. At the left
hand side of the condensate the density profile of positive particles is
close to one while small on the right hand side. The same reasoning holds
for the negative particles, we just have to exchange the word right with
left. Hence it is clear that the condensate as a whole moves diffusively
on the ring and therefore translational invariance is not broken.
Finite-size effects are much more important in the mixed phase than in the
pure phase. In the thermodynamical limit, in the condensate, the densities
$(+)$ and $(-)$ particles are both equal to $1/2$, hence the condensate can be
seen as a two species system with open boundaries in the ``maximal current
phase''. In the ``maximal current phase'', the current has the expression
\begin{eqnarray}
J=\frac{q-1}{4}
\label{eqI.VII}
\end{eqnarray}
which is precisely the value of the current observed in the whole mixed
phase. The condensate is glued to the fluid where all three species are
present and distributed in a uniform way. If we let now vary $q$ from 1 to
$q_c$, keeping $\lambda$ and $\rho$ constant, the condensate shrinks and the fluid
takes over the whole ring at $q_c$. In the whole mixed phase one has charge
segregation since the limit (\ref{eqI.IV}) exists. This brings us to two types of
charge segregation on the ring. One in which translational invariance is
broken (type B), like in the pure phase and another one in which 
translational invariance is unbroken (type UB), like in the mixed phase.

The {\em disordered phase} occurs when $q>q_c$, the current has a
non trivial dependence on the density.
The density profiles are uniform and there is no charge 
segregation.

\vspace{3mm}

  A lot of work was done previously on the three species diffusion problem
in other models, especially in two dimensions (see for example Ref.\cite{citI} and
references included). The present model is interesting for two reasons.
First, this model has the phase structure we just mentioned, secondly,
the stationary states of this model can be studied using the powerful
method of quadratic algebras. This not only allows to obtain precise
results but also allows the introduction of computational methods, like
the grand canonical ensemble which can not be defined otherwise. This
forces us to present our results starting with some mathematics.

  In Sec.\ref{secII} we first review some results obtained in Ref.\cite{citC} where it was
shown that for the rates (\ref{eqI.I}), the probability distribution for the
stationary state can be obtained by taking  traces of monomials of generators 
of a specific quadratic algebra. We give the representations of this
algebra. They are in general infinite-dimensional unless $\lambda$ is
related to $q$. We next show how to do calculations in the canonical
ensemble and explain why this approach can be applied only to small
lattices. A grand canonical ensemble is defined using the algebra. It is stressed
that this definition is not unique but very useful. We also show a novel
application of the algebra which gives the large $q$ (small $r$)
expansions of different physical quantities. A connection between the
representations of the quadratic algebra and the quantum algebra
$U_qso(2,1)$ is given in \ref{appA}.

  In Sec.\ref{secIII} we give the values of the currents in the three phases
using mainly Monte-Carlo simulations for the
canonical ensemble and algebraic methods for the grand canonical ensemble.
We also discuss the problem of fine-size corrections.

  The two-point correlation functions for the pure and mixed phase are
presented in Secs.\ref{secIV} and \ref{secV}. 

  Mean-field calculations are presented in \ref{appMF} and in Sec.\ref{secIMF} first
on the lattice and then in the continuum. The latter are interesting on
their own because we obtain stationary solutions of two coupled
non-linear differential equations in which the size $L$ of the system is a
parameter. The mean-field results are compared with various
results obtained through other methods.

  In Sec.\ref{secVI} we define an complex order parameter that we denote by $M$ and
compute the \FEF\  defined in Ref.\cite{citU} (there it was
called free energy functional) as a function of $|M|$. A minimum at a non-zero 
value of $|M|$ of the \FEF\ would imply charge
segregation. This is indeed seen for both the pure and mixed phase but
not, as expected, in the disordered phase. If we look however for a
system of size $L$ at the average time $T$ to move from a domain in the
complex $M$-plane to another, we observe that $T$ depends exponentially on $L$
in the pure phase and algebraically in the mixed phase. This implies
that translational invariance is spontaneously broken in the pure phase
but not in the mixed phase.

  At $q_c$ a second-order phase transition takes place. The values of several
critical exponents are given in Sec.\ref{secVII}. It is also shown that, like in the case
of Bose-Einstein condensation (reviewed in \ref{appBEC}), at $q_c$ the 
fugacities do not determine the densities.

  A detailed study of the phenomenon of spatial condensation is given in
Sec.\ref{secX}. We show the $L$ dependence of the compressibility and the
\FEF\ as a function of the local density of vacancies.
We also show that, amazingly, one can define a quantity which behaves like
the pressure in equilibrium systems. Based on what is done in the case of
Bose-Einstein condensation, we give a way to eliminate the condensate and
keep only the fluid in the mixed phase.

  Our conclusions are given in Sec.\ref{seccon}.

\section{The quadratic algebra and its applications.
\label{secII}}

In this section we show how to use the algebraic methods, pioneered by
Affleck et al \cite{citM} 
for equilibrium problems and by Hakim and Nadal \cite{citN} and
Derrida et al \cite{citO} for non-equilibrium problems, to describe the
stationary probability distribution obtained from the master equation
given by the rates (\ref{eqI.I}).

In Ref.\cite{citC} it was shown for which rates 
$g_{\alpha\beta}$ the 
unnormalized stationary probability
distribution $P_s(\beta_1,\beta_2,\ldots,\beta_L)$
of a three-state problem with $L$ sites on a ring
is given by the expression:
\begin{eqnarray}
P_s(\beta_1,\beta_2,\ldots,\beta_L)=\Tr (D_{\beta_1} D_{\beta_2} \cdots D_{\beta_L})
\qquad \qquad (\beta_k=0,1,2)
\label{eqII.I}
\end{eqnarray}
where the $D_\alpha$'s satisfy the quadratic algebra:
\begin{equation}
g_{\alpha\beta} D_{\alpha} D_{\beta} -g_{\beta \alpha} D_{\beta} D_{\alpha}
= x_{\beta} D_{\alpha} -x_{\alpha} D_{\beta}
\label{eqII.II}
\end{equation}

The cases for which the algebra (\ref{eqII.II}) exists and has representations with
a finite trace can be classified depending on how many of the parameters $x_\alpha$ are zero.
In particular, it was shown that for the rates considered in this paper,
the probability distribution can be obtained by the ansatz (\ref{eqII.I})--(\ref{eqII.II}) 
with $x_0=0$ (See Ref. \cite{citC}, Eqs.(3.30)--(3.47)). In this
case, the quadratic algebra takes the simple form:
\begin{eqnarray}
G_1 G_2 = r (G_2 G_1 + \lambda (G_1 + G_2) )		\nonumber\\
G_1 G_0 = G_0						\nonumber\\
G_0 G_2 = G_0					
\label{eqII.III}
\end{eqnarray}
where
\begin{eqnarray}
D_0=d_0 G_0\; , \qquad 
D_1=d_1 G_1\; , \qquad
D_2=d_2 G_2
\label{eqII.IV}
\end{eqnarray}
$d_0$,$d_1$ and $d_2$ are free parameters
(the freedom in choosing $x_1$ and $x_2$ was taken into account). 
The existence of the free parameters
is a consequence of the conservation of the numbers of particles $N_1$ and $N_2$.

We will first review the representations of the algebra (\ref{eqII.III}) obtained
in Ref. \cite{citC} and then explain how they can be used to compute various
physical quantities. 
We will also show
how to use the algebra (\ref{eqII.III}) to obtain a large $q$ (small $r$) 
expansion and
derive expressions for the current.
It turns out that in some cases, $G_1$ and $G_2$ can be
expressed in terms of generators of the quantum algebra $U_q(so(2,1))$. In
Appendix A we show how this information can be used. 

\subsection{Representations of the quadratic algebra.
\label{secII_I}}

As discussed already in Ref.\cite{citC}, we are interested for a given value
of $\lambda$ and $r$ in 
representations of the smallest
dimension.

A one-dimensional representation (this is equivalent to a product measure probability distribution)
is obtained for any $\lambda$ and $r$:
\begin{eqnarray}
G_0=0\; , \qquad
G_1=\gamma_1 \; , \qquad
G_2=\gamma_2
\label{eqII.V}
\end{eqnarray}
with
\begin{eqnarray}
\gamma_2=\frac{r \lambda \gamma_1}{\gamma_1 (1-r) -\lambda r}
\label{eqII.VI}\; .
\end{eqnarray}
Here $\gamma_1$ and $\gamma_2$ are c-numbers.
Since $G_0=0$, one has no vacancies and one has only an asymmetric
exclusion process. 

If $G_0$ is non-vanishing it is convenient to denote
\begin{eqnarray}
G_1 = {\cal A} + {\cal U}		\; , \qquad 
G_2 = {\cal A} + {\cal V}
\label{eqII.VII}
\end{eqnarray}
and a representation is given by:
\begin{eqnarray}
{G_0}= \left(\begin{array}{lllll}
		1&0&0&0&\cdots		\\	
		0&0&0&0&		\\
		0&0&0&0&		\\
		0&0&0&0&		\\
		\vdots&&&&\ddots	
	    \end{array}\right)		\;, \qquad
{\cal A}= \left(\begin{array}{lllll}
                a_1&0&0&0&\cdots    	\\
                0&a_2&0&0&         	\\
                0&0&a_3&0&		\\
		0&0&0&a_4&             	\\
                \vdots&&&&\ddots
            \end{array}\right)		\nonumber\\
{\cal U}= \left(\begin{array}{lllll}
                0&t_1&0&0&\cdots    	\\
                0&0&t_2&0&      	\\
                0&0&0&t_3&	        \\
                0&0&0&0&\ddots	        \\
                \vdots&&&&\ddots
            \end{array}\right)          \:, \qquad
{\cal V}= \left(\begin{array}{lllll}
                0&0&0&0&\cdots    	\\
                s_1&0&0&0&      	\\
                0&s_2&0&0&	        \\
                0&0&s_3&0&		\\
                \vdots&&&\ddots&\ddots
            \end{array}\right)
\label{eqII.VIII}
\end{eqnarray}
with
\begin{eqnarray}
a_k	=r\left[ (\lambda+1) \{k-1\}_r - \{k-2\}_r \right] \nonumber\\
s_k\, t_k=\{k\}_r\, \left[ (2\lambda+1)r-1+(r(\lambda+1)-1)^2 \,\{k-1\}_r \right]
\label{eqII.IX}\; .
\end{eqnarray}
We have used the notation
\begin{eqnarray}
\{k\}_r=\frac{r^k-1}{r-1}
\label{eqII.X}\; .
\end{eqnarray}

The representations given by Eq.(\ref{eqII.VIII}) 
are not necessarily infinite
dimensional. One can obtain an $n$-dimensional representation by finding the
solutions of the equation:
\begin{eqnarray}
s_n\, t_n = 0
\label{eqII.XI}\; .
\end{eqnarray}
Using Eqs.(\ref{eqII.IX}) and (\ref{eqII.XI}) we obtain:
\begin{eqnarray}
\lambda=\frac{1-r}{r(1+r^{\frac{1-n}{2}})}
\label{eqII.XII}
\end{eqnarray}
The $n$-dimensional representation has the following matrix elements ($k=1,2,\dots,n$)
\begin{eqnarray}
a_k     =\frac{1+r^{k-\frac{1+n}{2}}}{1+r^\frac{1-n}{2}} \nonumber\\
s_k\, t_k=\frac{(1-r^k)(1-r^{k-n})}{{\left(1+r^\frac{1-n}{2}\right)}^2}
\label{eqII.XIII}\; .
\end{eqnarray}
Let us observe that from Eq.(\ref{eqII.XII}) we get a one-dimensional
representation, $n=1$, 
with $G_0=G_1=G_2=1$ if
\begin{eqnarray}
q=2\lambda+1\,.
\label{eqII.XIV}
\end{eqnarray}
Notice that in this is case $G_0$ is non-zero!
A two-dimensional representation, $n=2$, is obtained if
\begin{eqnarray}
q=(1+\lambda)^2 
\label{eqII.XV}
\end{eqnarray}
and one can choose the basis such that one has
\begin{eqnarray}
\fl
{\cal A}= \left(\begin{array}{cc}
                1&0			\\
                0&\sqrt{r}		
            \end{array}\right)          \;, \quad
{\cal U}= \left(\begin{array}{cc}
                0&1-\sqrt{r}	        \\
                0&0
            \end{array}\right)          \;, \quad
{\cal V}= \left(\begin{array}{cc}
                0&0			\\
                \sqrt{r}-1&0
            \end{array}\right)\,.
\label{eqII.XVI}
\end{eqnarray}
A second two-dimensional representation is also obtained for any $\lambda$ if $r=0$. In this
case one has to take $r=0$ in the expressions for ${\cal A}$, ${\cal U}$ and 
${\cal V}$ in 
Eq.(\ref{eqII.XVI}).
This observation will be useful in Sec.\ref{secII_IV}.

To sum up, we have a one-dimensional representation with $G_0=0$ for any
$\lambda$ and $q$ and a second one-dimensional representation with $G_0\neq 0$ 
when $q$ and $\lambda$ are related by Eq.(\ref{eqII.XIV}). We also have two 
two-dimensional representations. One corresponds to Eq.(\ref{eqII.XV}), the second
one is obtained for $r=0$. For $n\geq 3$, one has only one $n$-dimensional
representation when Eq.(\ref{eqII.XII}) is satisfied. For generic $q$ and $\lambda$ and
$G_0\neq 0$ the representations are infinite-dimensional.

Due to the special structure of the representations with trace of the
algebra (\ref{eqII.III}), the expression (\ref{eqII.I}) has to be used with care. If one looks
at configurations with no vacancies, one has to use the one-dimensional
representation (\ref{eqII.V}). For any configuration which contains at least one
vacancy one has to use the representation given by Eqs.(\ref{eqII.VII})-(\ref{eqII.VIII}) which,
as discussed, can be finite or infinite-dimensional depending on the
values of $q$ and $\lambda$.

\subsection{The canonical ensemble.
\label{secII_II}}

In this case the numbers of particles $N_1$ and $N_2$ are given. The 
normalized probability distribution can be obtained from Eqs.(\ref{eqII.I}) and 
(\ref{eqII.IV}): 
\begin{eqnarray}
P^{\rm c}_{N_1,N_2}(\beta_1,\beta_2,\ldots,\beta_L)=
\frac{1}{Z_L^{N_1,N_2}} F^{\rm c}_{N_1,N_2}(\beta_1,\beta_2,\ldots,\beta_L)
\label{eqII.XVII}\; ,
\end{eqnarray}
where $Z_L^{N_1,N_2}$ is a normalization factor and
\begin{eqnarray}
F^{\rm c}_{N_1,N_2}(\beta_1,\beta_2,\ldots,\beta_L)=
\tr (G_{\beta_1}\,G_{\beta_2}\,\cdots G_{\beta_L}) \nonumber\\
\qquad\times \delta(\sum_{k=1}^L \delta(\beta_k-1)-N_1)\,
\delta(\sum_{k=1}^L \delta(\beta_k-2)-N_2)
\label{eqII.XVIII}
\end{eqnarray}
where $\delta(x)$ is the Kronecker function.

The currents
are going to play an important role in this paper. They are defined
as follows:
The current for the particles of type 1, 
taken for convenience on the link between the sites
$L-1$ and $L$, is:
\begin{eqnarray}
\fl 
j_1=\lambda\, \delta(\beta_{L-1}-1) \, \delta(\beta_L)
+ q\, \delta(\beta_{L-1}-1) \, \delta(\beta_L-2)
- \delta(\beta_{L-1}-2) \, \delta(\beta_L-1)
\label{eqII.XIX}
\end{eqnarray}
The current for the type 2 particles has a similar expression. Using 
the Eqs.(\ref{eqII.XVII})-(\ref{eqII.XIX}) and (\ref{eqII.III}), 
the expression for the average current
is:
\begin{eqnarray}
\fl
J_1=\frac{\lambda}{Z_L^{N_1,N_2}} \sum_{\beta_1,\ldots,\beta_{L-2}}
[ \tr (G_{\beta_1}\,G_{\beta_2}\,\cdots G_{\beta_{L-2}}\, G_0 )\nonumber\\
\qquad\qquad\times\delta(\sum_{k=1}^{L-2} \delta(\beta_k-1)-N_1+1)\,
\delta(\sum_{k=1}^{L-2} \delta(\beta_k-2)-N_2)  \nonumber\\
\qquad + \tr (G_{\beta_1}\,G_{\beta_2}\,\cdots G_{\beta_{L-2}}\, (G_1 + G_2))\nonumber\\
\qquad\qquad\times\delta(\sum_{k=1}^{L-2} \delta(\beta_k-1)-N_1+1)\,
\delta(\sum_{k=1}^{L-2} \delta(\beta_k-2)-N_2+1) ]
\label{eqII.XX}
\end{eqnarray}
It is useful to employ the Fourier transform of the Kronecker delta function
\begin{eqnarray}
\delta(\alpha)=\frac{1}{L} \sum_{k=0}^{L-1} \exp(\frac{2\pi {\rm i}}{L}k \alpha)
\qquad\qquad (\alpha=0,\ldots,L-1)
\label{eqII.XXI}
\end{eqnarray}
and to define
\begin{eqnarray}
B^{(L)}_{r,s}=G_0+\exp(\frac{2\pi{\rm i}}{L} r)G_1+\exp(\frac{2\pi{\rm i}}{L} s)G_2
\label{eqII.XXII}\; .
\end{eqnarray}
In this way one can get a compact expression for the current of type 1 
particles. The expressions for the type 2 particles and for the 
normalisation factor $Z^{N_1,N_2}_L$ 
can be obtained in a similar way. One obtains:
\begin{eqnarray}
\fl
Z_L^{N_1,N_2}=\frac{1}{L^2} \sum_{r,s=0}^{L-1}
	\exp\left(-2\pi{\rm i}\left(r\frac{N_1}{L}+s\frac{N_2}{L}\right)\right) \tr \left( (B^{(L)}_{r,s})^L \right)
\label{eqII.XXIII}\\
\fl
J_1=\frac{\lambda}{Z_L^{N_1,N_2}} \sum_{r,s=0}^{L-3}
	\exp\left(-2\pi{\rm i}\left(r\frac{N_1-1}{L-2}+s\frac{N_2}{L-2}\right)\right) 
	\tr \left( (B^{(L-2)}_{r,s})^{L-2} B^{(L-2)}_{s,s} \right)
\label{eqII.XXIV}\\
\fl
J_2=\frac{\lambda}{Z_L^{N_1,N_2}} \sum_{r,s=0}^{L-3}
	\exp\left(-2\pi{\rm i}\left(r\frac{N_1}{L-2}+s\frac{N_2-1}{L-2}\right)\right)
	\tr \left( (B^{(L-2)}_{r,s})^{L-2} B^{(L-2)}_{r,r} \right)
\label{eqII.XXV}
\end{eqnarray}
Analogous expressions can be written for the correlation functions.

In a concrete calculation ($q$ and $\lambda$ are given) we take  
$G_0$, $G_1$ and $G_2$ from 
Eqs.(\ref{eqII.VIII}) and (\ref{eqII.IX}). 
This gives the matrices $B_{r,s}^{(L)}$
of Eq.(\ref{eqII.XXII}) and we are left with the calculation of the traces appearing in
Eqs.(\ref{eqII.XXIII})-(\ref{eqII.XXV}) 
and the double summations over $r$ and $s$. We have been 
unable to do any analytical calculation in this way in the general 
case (for finite dimensional representations this is probably 
possible). 
This takes us to numerical calculations. If we take $N_1+N_2 \neq L$, the trace 
operation in Eq.(\ref{eqII.XX}) is well defined and any divergence in Eqs. 
(\ref{eqII.XXIII})-(\ref{eqII.XXV}) has to cancel, 
this does not make however numerical evaluations 
easy.  
To circumvent the problem of computing traces of infinite dimensional
matrices 
we will assume that we have at least one vacancy and
take in Eq.(\ref{eqII.XXIII}) $\tr ( G_0 (B^{(L-1)}_{r,s})^{L-1} )\equiv
\{(B^{(L-1)}_{r,s})^{L-1} \}_{1,1}$
instead of $\tr ( (B^{(L)}_{r,s})^L )$ and $L-1$ instead of $L$.
The expressions for the currents have to changed in the same sense.
Due to the fact that $G_0$ is a projector (see Eq. (\ref{eqII.VIII}))
this procedure has also the advantage that for a 
problem with $L$ sites, only the matrix elements with $k<L$ (see Eq.(\ref{eqII.VIII})) appear in 
the evaluation of the traces.
In this way we have been able to compute the current up to
$L=30$ with high precision.
This allows in principle, using extrapolants \cite{citP},
to obtain the current in the thermodynamical limit. As will be shown in
Sec.\ref{secIII} the current might converge very slowly towards its thermodynamical
limit and therefore in these cases this method is useless. As a matter of
fact, throughout this paper, we have used Monte-Carlo simulations for 
the canonical ensemble. The precision is much worse than using
the analytical method but one can reach much larger values of $L$.

\subsection{The grand canonical ensemble.
\label{secII_III}}

As opposed to equilibrium states, the definition of a grand canonical 
ensemble for stationary states is not unique. In principle, for the 
master equation one can give, at $t=0$, an initial probability 
distribution in which the number of particles $N_1$ and 
$N_2$ are not fixed 
but the ``chemical potentials'' $\mu_1$ 
and $\mu_2$ are. Such an initial 
probability distribution has the expression:
\begin{eqnarray}
P^{gc}_{\mu_1,\mu_2}(\beta_1,\ldots,\beta_L,t=0)=
\sum_{N_1,N_2} 
  Q(\beta_1,\ldots,\beta_L;N_1,N_2) e^{\mu_1 N_1+\mu_2 N_2}
\label{eqII.XXVI}
\end{eqnarray}
where $Q$ is an arbitrary function.
This initial condition determines a certain stationary probability
that can be considered as a grand canonical ensemble.
The $N_1$ and $N_2$ dependence of the functions $Q$ 
remains however arbitrary even if 
we make sure that the average densities $\rho_1=<\!N_1\!>/L$ and 
$\rho_2=<\!N_2\!>/L$ are properly obtained. We 
will adopt a different 
approach looking directly at the canonical 
stationary distribution given by Eq.(\ref{eqII.XVII}) 
and consider 
the grand 
canonical ensemble defined by the expression:
\begin{eqnarray}
P^{gc}_{\mu_1,\mu_2}(\beta_1,\ldots,\beta_L)=
\frac{1}{Z_L^{\mu_1,\mu_2}}
\sum_{N_1,N_2} 
\frac{F^{\rm c}_{N_1,N_2}(\beta_1,\beta_2,\ldots,\beta_L)}{f^L_{N_1,N_2}}
e^{\mu_1 N_1+\mu_2 N_2}
\label{eqII.XXVII}
\end{eqnarray}
where we have used Eq.(\ref{eqII.XVIII}). 
In Eq.(\ref{eqII.XXVII}) the factors $f^L_{N1,N2}$ are 
arbitrary. We will take from now on:
\begin{eqnarray}
f^L_{N_1,N_2}=1
\label{eqII.XXVIII}
\end{eqnarray}
A similar definition was used in \cite{DeJaLeSp}.
This gives:
\begin{eqnarray}
P^{gc}_{\mu_1,\mu_2}(\beta_1,\ldots,\beta_L)=
\frac{1}{Z_L^{\mu_1,\mu_2}}
\tr (E_{\beta_1}\cdots E_{\beta_L})
\label{eqII.XXIX}
\end{eqnarray}
where
\begin{eqnarray}
E_0=G_0,\quad\quad
E_1=z_1 G_1,\quad\quad
E_2=z_2 G_2
\label{eqII.XXX}
\end{eqnarray}
and 
\begin{eqnarray}
z_1=e^{\mu_1},\quad\quad z_2=e^{\mu_2}
\label{eqII.XXXI}
\end{eqnarray}
are the fugacities.
As opposed to the canonical ensemble, the grand canonical partition 
function has a simple expression:
\begin{eqnarray}
Z_L^{\mu_1,\mu_2}=\tr \left(C^L\right)
\label{eqII.XXXII}
\end{eqnarray}
where
\begin{eqnarray}
C=E_0+E_1+E_2
\label{eqII.XXXIII}\; .
\end{eqnarray}
That this choice for the definition of the grand canonical ensemble is 
not unique can be seen in a different way. 
Instead of using the $D$'s of 
Eq.(\ref{eqII.IV}) with $d_0=d_1=d_2=1$
(for the canonical ensemble 
this choice is not important), 
we could have taken $d_0=1$ (one of the $d_i$
can be chosen at will) but keep $d_1$ and 
$d_2$ as free parameters. In this 
case, Eq.(\ref{eqII.XXX}) is replaced by 
\begin{eqnarray}
E_0=G_0,\quad\quad
E_1=d_1 z_1 G_1,\quad\quad
E_2=d_2 z_2 G_2
\label{eqII.XXXIV}
\end{eqnarray}
and all the results for finite-size systems will be affected, but hopefully 
not the results for the thermodynamical limit. We want to stress this 
point because the interpretation of finite-size corrections for the 
grand canonical ensemble has still to be understood. We have 
arbitrarily taken the choice given by Eq.(\ref{eqII.XXVIII}). 
Our choice promotes the algebra (\ref{eqII.III})
to a special role,
since the isomorphic algebras obtained using various values of the $d_i$ in Eq.(\ref{eqII.IV})
give different definitions of the grand canonical ensemble.
We have no good reason to prefer one algebra to the other except the fact
that the representation of the algebra (\ref{eqII.III})  
can be connected ``naturally'' to known representations
of quantum algebras (see Appendix A).

The expression (\ref{eqII.XXIX}) reminds us of an analogous expression
in equilibrium statistical physics and along this line of thought,
one could perhaps find a better justification for
a proper definition of the grand canonical ensemble.
That this idea is not too far fetched will be seen in Sec.\ref{secX}.

We now proceed with the description of the grand canonical ensemble. 
As usual the fugacities are fixed by the densities: 
\begin{eqnarray}
\rho_i=\frac{<N_i>}{L}=
\frac{z_i}{L} \frac{\partial\log Z_L^{\mu_1,\mu_2}}{\partial z_i}
\label{eqII.XXXV}
\end{eqnarray}
The fluctuations of the densities are:
\begin{eqnarray}
<(\Delta N_i)^2>=\frac{\partial^2 \log Z_L^{\mu_1,\mu_2}}
			{{\partial\log z_i}^2}
\label{eqII.XXXVI}
\end{eqnarray}
Using Eqs.(\ref{eqII.XXIX}) and 
(\ref{eqII.XXXIII}), one gets the two-point functions:
\begin{eqnarray}
c_{\alpha,\beta}(R)=\frac{1}{Z_L^{\mu_1,\mu_2}}
    \tr(G_\alpha C^{R-1} G_{\beta} C^{L-R-1})
\label{eqII.XXXVII}
\end{eqnarray}
Notice that if the matrix $C$ 
has the two
largest eigenvalues $\lambda_1>\lambda_2$ and if 
these eigenvalues are not 
degenerate, at large distances $R$ the two point 
function has the expression:
\begin{eqnarray}
c_{\alpha,\beta}(R) \propto \exp (-MR)
\label{eqII.XXXVIII}
\end{eqnarray}
where the inverse correlation length $M$ is:
\begin{eqnarray}
M=\log \frac{\lambda_1}{\lambda_2}
\label{eqII.XXXIX}
\end{eqnarray}
This implies that for finite-dimensional representations 
(see Eq.(\ref{eqII.XIII}))
one expects finite correlation lengths. This conjecture can be proven 
rigorously but it is a lengthy exercise that we have not done.
The expression of the currents can be derived using 
Eqs.(\ref{eqII.XIX})
and (\ref{eqII.III}). We get
\begin{eqnarray}
J_1=\frac{\lambda z_1}{Z_L^{\mu_1,\mu_2}} 
\tr\left((G_0+z_2(G_1+G_2)) C^{L-2}\right)
\label{eqII.XXXX}\; .
\end{eqnarray}
The expression for $J_2$ is obtained by permuting 1 with 2 in 
Eq.(\ref{eqII.XXXX}).
If we consider the case in which the densities of the particles 1 and 2 
are the same, the expression of the currents takes a very simple form:
\begin{eqnarray}
J_1=J_2=\lambda z \frac{Z_{L-1}^{\mu,\mu}}{Z_{L}^{\mu,\mu}}
\label{eqII.XXXXI}
\end{eqnarray}
where
\begin{eqnarray}
z=e^{\mu}
\label{eqII.XXXXII}\; .
\end{eqnarray}
This makes the calculation of the current much simpler in the grand 
canonical ensemble.

In the expressions written above (see (\ref{eqII.XXXII}),
(\ref{eqII.XXXVII}) and (\ref{eqII.XXXX})), 
we 
have made the assumption that the trace operation exists. This is 
clearly the case if one has finite-dimensional representations. 
The case 
of infinite-dimensional representations has to be considered with care. 
In the grand canonical ensemble, for a given number of sites $L$ 
one sums 
over the values $N_1$ and $N_2$. 
In this sum there is also the term with
$N_1+N_2=L$ where no $G_0$ matrix appears (\ref{eqII.XVIII}) 
and one has only products of $G_1$ and $G_2$ 
matrices given by Eq.(\ref{eqII.VIII}) which have no trace. 
One has
to use for this case the representation 
(\ref{eqII.V}) which makes the calculation cumbersome. 
A simpler way 
is to define the grand canonical ensemble such that one has at least 
one vacancy. That is what we have done.
In this case, one has to sum over all possibilities for the relative
position of the $G_0$ matrix and the 
operator whose expectation value one wants to compute
(see Eqs.(\ref{eqII.XXXII}),(\ref{eqII.XXXVII}) and (\ref{eqII.XXXX})). 
For each term, taking into 
account the form of the matrix $G_0$ (see Eq.(\ref{eqII.VIII})), 
one replaces the 
trace operation by taking an appropriate $(1,1)$ matrix element.
%
%
%
%
Like in the canonical ensemble, we have not been able to do analytical 
calculations and had to use the computer but in the grand canonical 
ensemble we were able to reach 
$L=300$,
as opposed to $L=30$ for the canonical ensemble.

\subsection{The large $q$ expansion in the grand canonical ensemble.
\label{secII_IV}}

A useful application of the algebra (\ref{eqII.III}) is that one can do a large $q$ 
(small $r$) expansion in an algebraic way.
We will show how to obtain the physical quantities up to order $r$.
We first note that, if we keep terms
up to order $r$ in the matrix elements (\ref{eqII.IX}) which give $G_1$ and $G_2$, the 
only non-vanishing elements are
\begin{eqnarray}
a_1=1\; ,\qquad
a_2=(1+\lambda)r\; ,\qquad
a_k=\lambda r\quad (k\geq 3)\nonumber\\
s_1 t_1 = (2\lambda +1)r -1\; ,\qquad s_2 t_2 = -r
\label{eqII.XXXXIV}
\end{eqnarray}
Since one is interested in the smallest representation, it is enough to 
take a $3\times 3$ dimensional one. One can check that taking this 
representation one obtains:
\begin{equation}
G_1 G_2 - r \left(G_2 G_1 + \lambda (G_1+G_2)\right)=
r^{\frac32}\left(\begin{array}{ccc}
		0&0&0\\
		0&0&-(1+\lambda)\\
		0&(1+\lambda)&0
		\end{array}\right)
\label{eqII.XXXXV}
\end{equation}
which shows the consistency of the approach since the right-hand-side of Eq. (\ref{eqII.XXXXV}) is of order $r^{3/2}$. 
We have chosen a representation in which ${\cal U}=-{\cal V}^T$
(see Eq.(\ref{eqII.VIII})).
Notice that
if we neglect the terms of 
order $r$, we obtain the two dimensional representation (\ref{eqII.XVI}) in which 
we take $r=0$.

We will consider only the grand canonical ensemble and assume that 
the densities of particles 1 and 2 are the same:
\begin{equation}
\rho=\rho_1=\rho_2
\label{eqII.XXXXVI}
\end{equation}
The matrix $C$ given by Eq.(\ref{eqII.XXXII}) becomes
\begin{equation}
C=\left(\begin{array}{ccc}
	1+z_1+z_2 & z_1 (1-\frac{2\lambda+1}{2}r) & 0 \\
	-z_2 (1-\frac{2\lambda+1}{2}r) & (1+\lambda)(z_1+z_2)r & z_1 r^{\frac{1}{2}} \\
	0 & -z_1 r^{\frac{1}{2}} & \lambda (z_1+z_2) r
	\end{array}\right)
\label{eqII.XXXXVII}\; .
\end{equation}

We are interested to obtain the coefficients of the $r$-expansion defined 
by the expressions:
\begin{eqnarray}
\lambda_1(z_1,z_2)=\lambda_1^{(0)}(z_1,z_2) + r \lambda_1^{(1)}(z_1,z_2) 	\nonumber\\
z_{1/2}=z^{(0)}_{1/2}(\rho)+ r z^{(1)}_{1/2} (\rho)						\nonumber\\
J_1=J_1^{(0)}(\rho) + r J_1^{(1)}(\rho)
\label{eqII.XXXXVIII}
\end{eqnarray}
where $\lambda_1(z_1,z_2)$ is the largest eigenvalue of  $C$.
We didn't look for higher order terms, which would imply taking 
larger matrices than $3\times 3$.

We start with the leading order in which case we have to deal with a 
$2\times 2$ matrix. The two eigenvalues of the matrix $C$ are:
\begin{eqnarray}
\lambda^{(0)}_{1/2}=\frac{1}{2}\left( 1+z_1+z_2\pm\sqrt{(1+z_1+z_2)^2-4z_1 z_2} \,\right)
\label{eqII.XXXXIX}
\end{eqnarray}
Using Eqs.(\ref{eqII.XXXII}) and (\ref{eqII.XXXV}) we get:
\begin{equation}
z^{(0)}_{1/2}(\rho)=\frac{\rho(1-\rho)}{(1-2\rho)^2}
\label{eqII.XXXXX}
\end{equation}
Introducing the expression (\ref{eqII.XXXXX}) in 
Eq.(\ref{eqII.XXXXIX}) we find the eigenvalues 
of the matrix $C$ at a given density:
\begin{eqnarray}
\lambda_1^{(0)}(z,z)=\left(\frac{1-\rho}{1-2\rho}\right)^2 
\nonumber\\
\lambda_2^{(0)}(z,z)=\left(\frac{\rho}{1-2\rho}\right)^2 
\label{eqII.XXXXXII}
\end{eqnarray}
Using now Eq.(\ref{eqII.XXXIX}), we derive in zero-th order the expression of the 
inverse correlation length:
\begin{equation}
M=2 \log\left(\frac{1-\rho}{\rho}\right)
\label{eqII.XXXXXIII}
\end{equation}
It is interesting to observe that the ``mass'' vanishes for the maximum 
value of the density ($\rho=1/2$) and diverges at small densities.
We can now use the expression of the partition function 
\begin{eqnarray}
Z_L^{\mu,\mu}=\left(\lambda_1^{(0)}\right)^L
\label{eqII.XXXXXIV}
\end{eqnarray}
and Eq.(\ref{eqII.XXXXI}) to get the expression of the current:
\begin{equation}
J_1^{(0)}=\frac{\lambda\rho}{1-\rho}
\label{eqII.XXXXXV}
\end{equation}
To get the next order terms in $r$, we use the expression (\ref{eqII.XXXXVII})
for the matrix $C$ and get the correction to the largest eigenvalue. We 
obtain:
\begin{equation}
\lambda_1^{(1)}(z_1,z_2)=\frac{\alpha\left(\lambda_1^{(0)}\right)^2
				-\beta\lambda_1^{(0)} +\gamma }
			      {3 \left(\lambda_1^{(0)}\right)^3
				-2 \delta \lambda_1^{(0)} +z_1 z_2 }
\label{eqII.XXXXXVI}
\end{equation}
where
\begin{eqnarray}
\alpha=(1+2\lambda)(\delta-1)				\nonumber\\
\beta=(1+2\lambda)\delta(\delta-1) - 2 \lambda z_1 z_2		\nonumber\\
\gamma=\left( (1+\lambda) \delta -\lambda \right) z_1 z_2 	\nonumber\\
\delta=1+z_1+z_2
\label{eqII.XXXXXVII}
\end{eqnarray}
It is obvious how to proceed further. We get:
\begin{eqnarray}
\fl
\lambda_1^{(1)}(z,z)=\rho^2\frac{1+2\lambda -2\rho -2\lambda\rho}
			  {(1-2\rho)^2} 				\nonumber\\
\fl
z^{(1)}=-\rho^2\frac{\lambda +(1-3\rho +\rho^2)(1+\lambda)}
			  {(1-\rho)^2\, (2\rho^2 + 1 -2\rho)}		
\label{eqII.XXXXXVIII}\\
\fl
J_1^{(1)}=-\rho^2\lambda \frac{1+2\lambda -\rho(6+9\lambda)+\rho^2 (12+13\lambda)-\rho^3 (6-2\lambda)
		-\rho^4(6+8\lambda)+4\rho^5(1+\lambda) }
			  {(1-\rho)^4\, (2\rho^2 + 1 -2\rho)}		\nonumber
\end{eqnarray}
One interesting result that one can read from Eq.(\ref{eqII.XXXXXVIII}) 
is that, at 
fixed $\lambda$, the current reaches the asymptotic value for $q\rightarrow\infty$
from below.

\section{The currents. \label{secIII}}
\fII
From now on, the particles of type 1 and 2 will be called positive
respectively negative because we think that it is easier to visualise the
physics in this way. Their densities will be denoted by $p$ respectively
$m$ (the density of vacancies will be $v$). If the densities of positive and
negative particles are equal (this is going to be mostly the case), we
will write
\begin{eqnarray}
p=m=\rho\;.
\label{eqIII.I}
\end{eqnarray}

It is useful to give the expression of the currents in the mean-field
approximation. From Eq.(\ref{eqII.XIX}) we get:
\begin{eqnarray}
J_+=\lambda p v + (q-1) p m\nonumber\\
J_-=\lambda m v + (q-1) p m
\label{eqIII.II}
\end{eqnarray}
where obviously
\begin{eqnarray}
p+m+v=1\;.
\label{eqIII.III}
\end{eqnarray}
As is shown in \ref{appMF} (see also Sec.\ref{secIMF}), the mean-field equations
have not only homogeneous solutions. 
Here we refer to the homogeneous solutions.
We will give first some pictures showing the $q$ 
dependence of the currents at
fixed values of $\lambda$ and densities. These pictures are just a few from the
data we collected. Our conclusions will be based on all the available
information. The thermodynamical limit for the values of the current have
been obtained using the algebraic approach to the grand canonical ensemble
(see Sec.\ref{secII_III}) up to $L=200$. 
The large $L$ limit has been obtained using
extrapolants \cite{citP}. 
All the results have been checked using Monte-Carlo
simulations (this corresponds to the canonical ensemble).

Since we consider the case in which the densities of
positive and negative particles are the same and therefore one has:
\begin{eqnarray}
J_+=J_-=J
\label{eqIII.IV}
\end{eqnarray}
In Fig.\ref{fII} we show the current for fixed $\lambda=1$ and two values of
the density ($\rho=0.2$ and $0.4)$. 
The large $q$ behavior of the current is not shown in the picture since its expression
is known from Eq. (\ref{eqII.XXXXVIII}).
We distinguish three domains that for good
reasons we are going to call phases. The names for the phases will look
natural in the next sections.
\begin{itemize}
\item[a)]
{\em Pure phase} ($0<q<1$) where the current is zero.
\item[b)]
{\em Mixed phase} ($1<q<q_c$) where the current is given by
\begin{eqnarray}
J=\frac{q-1}{4}
\label{eqIII.V}
\end{eqnarray}
(see the line $(c)$ in Fig.\ref{fII}), for both densities. The critical value
of $q$ denoted by $q_c$ is density dependent.
\item[c)]
{\em Disordered phase} ($q>q_c$) where the current is density dependent. For
large values of $q$, the current is given by the large 
$q$-expansion (see
Eqs.(\ref{eqII.XXXXVIII}), (\ref{eqII.XXXXXV}) and (\ref{eqII.XXXXXVIII})).
\end{itemize}
The solid lines $(a)$ and $(b)$ in Fig.\ref{fII} 
are the mean-field values obtained from
Eq.(\ref{eqIII.II}). 
One can notice two facts. 
First, for $q=3$ mean-field gives
the current correctly. 
This should not come as a surprise since from
Eq.(\ref{eqII.XIV}) with $\lambda=1$, 
we learn that the quadratic algebra has
a one-dimensional representation, which implies that mean-field is exact.
Next, we observe that, for both densities,
mean-field gives correctly the currents for values of $q$ close to $q_c$.
This result has deeper consequences as will be seen in Sec.\ref{secIMF},
where it will be shown that at $q_c$ the current is given by mean-field
and that the discrepancy shown in Fig.1 comes probably from finite-size effects (one extrapolates from $L=200$).

\fIa
Having in mind Eq.(\ref{eqIII.V}), 
in Fig.\ref{fIa} 
we show the quantity $4J/(q-1)$ as a
function $q$ for fixed density ($\rho=0.2$) and 
three values of $\lambda$. 
The same three phases are present. We notice that the current vanishes in the
pure phase, that it is given by 
Eq.(\ref{eqIII.V}) in the mixed phase and that $q_c$
increases with $\lambda$. 
For $\lambda=0$ the current vanishes for any value of $q$
(see Eq.(\ref{eqII.XX})).

Similar calculations for other densities 
and values of 
$\lambda$ have
shown that in the pure phase the current vanishes and that in the mixed
phase it is given by the simple expression 
(\ref{eqIII.V}) independent of density
and $\lambda$. 
This is a very unexpected result. The
values of $q_c$ are dependent on 
$\lambda$ and on the density. 
In Sec.\ref{secIMF} we will derive the expressions of $q_c$ (see Eq.(\ref{eqV.XXVI}) for equal
densities) and the finite-size
corrections. The interested reader can already have a look at Fig.\ref{fC} where
the density dependence of the current is shown for various lattice sizes.

\fXIII
It is obvious from the Figs.\ref{fII} and \ref{fIa} 
that at $q_c$ the derivative of the
current has a discontinuity. In Fig.\ref{fXIII} 
we show other local quantities,
which are various connected correlations functions at distance 1 (one
lattice site) for $\lambda=1$ and 
$\rho=0.2$. The large $L$ limit was obtained using
the grand canonical ensemble extrapolating 
from data obtained up to $L=150$.
One notices that the derivatives of all these quantities have also a
discontinuity at $q_c=1.62$ (see Table \ref{tI}). 
The Fig.\ref{fXIII}
shows also an interesting
fact: 
for reasons which remain a mystery, the connected function $c_{-+}(1)$
practically vanishes in the disordered phase such that the discontinuity of its
derivative is easier to determine. 
This allows the best estimate of $q_c=1.62\pm 0.04$.

%
%
%
%
%
%
%
%
%

In the next sections we are going to discuss at length the three
phases. 
It is however very instructive to describe what one can see on the screen of the terminal while 
looking at the succession of configurations of a single Monte-Carlo 
run. 
This is of course a qualitative picture only.
First, let us consider the pure phase. 
In order to fix the 
ideas, assume $q=0.8$, $\lambda=1$ and $\rho=0.2$ (see Fig.\ref{fII}). 
For lattices lengths
$L<L_1$ ($L_1 \approx 20$), one sees a uniform distribution of positive, negative  particles
and vacancies. 
For $L_1<L<L_2$ ($L_2 \approx 30$) one observes a separation into two domains: 
one containing only positive and negative particles and one in which one 
has predominantly vacancies but also some charged particles. For $L>L_2$ one 
sees three domains one purely positive, one purely negative and one 
purely neutral (hence the name pure phase). 
These domains look like 
pinned in space (they move very slowly from one place on the ring to 
another one). 
In the mixed phase, (take $q=1.2$, $\lambda=1$, $\rho=0.2$, see again
Fig.\ref{fII}) the evolution with $L$ of the relevant configurations is similar but by 
no means identical. Again for $L<L_1$ ($L_1 \approx 80$) the distributions are uniform, 
For $L_1<L$ there is again a separation into two domains. 
In the first, one has
only positive and negative particles with non-uniform distributions, we 
call it {\em condensate}. 
In the second domain, positive, negative particles 
and vacancies are uniformly distributed, we call this domain {\em fluid}. 
Increasing L, the concentrations inside the condensate change slowly and 
get more uniform and the fluid survives as a domain with a uniform 
distribution of the three species. 
This is one major difference between 
the pure and mixed phases. 
A second one comes from the observation that
these profiles are not anymore pinned in the space but have a Brownian
motion on the ring.

Before closing this section we would like to discuss the problem
of finite-size corrections to the current.

It is amusing to start with a simple problem. Let us take 
$\lambda=1$ and
$q=3$ and equal densities $\rho$. 
As already observed before, in this case one
has a one-dimensional representation of the quadratic algebra (\ref{eqII.III}) 
which can
be obtained using Eq.(\ref{eqII.XIII}): 
\begin{eqnarray}
G_0=G_1=G_2=1
\label{eqIII.VIII}
\end{eqnarray}
One can then do the calculations in the canonical ensemble where one gets
in leading order
\begin{eqnarray}
J=\rho(1+\frac{1}{L})
\label{eqIII.IX}
\end{eqnarray}
and in the grand canonical ensemble where one obtains
\begin{eqnarray}
J=\rho
\label{eqIII.X}
\end{eqnarray}
for any number of sites. This result is surprising since the parameters
are such that we are in the disordered phase and one would expect exponential
corrections (as will be shown in Sec.\ref{secVII} the correlation length is finite
in this phase).

We have looked in several cases to the finite-size correction terms and
the results are given below. 
All data are for $\lambda=1$, $\rho=0.2$. First we
consider the grand canonical ensemble:
\begin{eqnarray}
q=0.8\qquad &J=0.09 \exp(-0.055 L)	\qquad&\mbox{(pure phase)}\nonumber\\
q=1.2\qquad &J=0.05 + 1.6/L	 	\qquad&\mbox{(mixed phase)}\nonumber\\
q=2.2\qquad &J=0.18158 + 0.04/L^2 	\qquad&\mbox{(disordered phase)}\nonumber\\
q=3.8\qquad &J=0.21065 - 0.016/L^2 	\qquad&\mbox{(disordered phase)}
\label{eqIII.XI}
\end{eqnarray}

\fIVa
For the canonical ensemble, it is harder to estimate the correction
terms. In order to illustrate the problem, we take $q=1.2$ in the mixed phase. 
In Fig.\ref{fIVa} 
we
show the deviation of the current from its asymptotic value (0.05) 
as a
function of the number of sites $L$. 
The Monte-Carlo data are shown
together with the results obtained using the approach presented in Sec.\ref{secII_II}.
Looking at the Monte-Carlo data first (up to $L=200$) 
one observes that one
gets large corrections. (The Monte-Carlo data are not precise enough to
use extrapolants). On the other hand, the data obtained in the algebraic
approach 
are in a range of $L$ 
in which any extrapolation is bound to give
wrong results. 
In Fig.\ref{fIVa} we also show the values of the current obtained from the inhomogeneous solutions of the 
mean-field equations (see Appendix B). 
At large values of $L$ the current obtained in this way  is in agreement with the Monte-Carlo
data, but at small values of $L$, they show a curious behavior.
The explanation of this phenomenon and a more extensive discussion
of the finite-size corrections in the mixed phase for the canonical ensemble can be found in Sec.\ref{secIMF}.
There is a case 
in which the finite-size corrections
converge very nicely. This is the case $q=1$. 
Moreover, in Sec.\ref{secIMF} we will
even show how to get them theoretically. Here, we just give the
Monte-Carlo results, again for $\lambda=1$ but for different densities:
\begin{eqnarray}
\rho=0.1\qquad&     J={5}/{L} 	\nonumber\\
\rho=0.2\qquad&     J={2.5}/{L}	\nonumber\\
\rho=0.4\qquad&     J={1.25}/{L}
\label{eqIII.XII}
\end{eqnarray}
The reader can easily guess the analytical expression of the correction term.
In order to help the reader, we have not given the errors in Eq. (\ref{eqIII.XII}) (they are small).

We will now discuss in detail the three phases.
\section{The pure phase ($0<q<1$) \label{secIV}}

\fXXXI

In the last section we have mentioned that in the pure phase the current vanishes
exponentially with the lattice size for any values of the densities (they might
be different) and of $\lambda$. 
We will now get a complete picture of this phase (it
is very simple).

At $q=0$, a single vacancy
is sufficient to break the
translational invariance of the system.
At a finite density of
vacancies, the ground state of the time evolution 
hamilton operator is infinitely degenerate 
for $L=\infty$, each configuration
of the kind
\begin{equation}
\fl
(\,0\,)\cdots (\,0\,)(+)\cdots (+)(-)\cdots (-)
(\,0\,)\cdots (\,0\,)(+)\cdots (+)(-)\cdots (-)\cdots
\label{eqIV.I}
\end{equation}
being a stationary state.
At $q$ different from zero and
finite $L$, if one starts with an arbitrary configuration, the system organizes
itself into only three blocks
\begin{equation}
\label{eqIV.II}
\fl
(\,0\,)\cdots (\,0\,)(+)\cdots (+)(-)\cdots (-)
\end{equation}
which cover the lattice.
Translational invariance is respected but as shown in Sec.\ref{secVI} the blocks hop from one
position to another with a flip time which increases
exponentially
with $L$. 
This implies that for $q<1$, translational invariance is
spontaneously broken. Since each of the three blocks contains one kind of
particle only, we call it the pure phase.

We postpone the discussion of the flip times
and of the behavior of the order-parameter which describes the breaking of
translational invariance. 
This is going to be done in Sec.\ref{secVI} 
in conjunction with a
similar problem but with different results for the mixed phase.

In Fig.\ref{fXXXI}
we show the two-point correlation functions 
$c_{0,0}$ and $c_{+,-}$ 
defined by Eq.(\ref{eqII.XXXVII}) not as a function of the distance 
$R$ measured in lattice
sites but of the macroscopic distance
\begin{eqnarray}
y=R/L  
\label{eqIV.III}
\end{eqnarray}

For any fixed $y$, 
the data converge exponentially to their thermodynamic
limit. This explains why the correlations functions for so small lattices 
($L=75$
and $100$) coincide already.
The expression of the correlation functions for very small lattices can be derived
from the mean-field calculations presented in Sec.\ref{secIMF}.
It was checked using Monte-Carlo simulations that these calculations give the correct picture of finite-size effects.

We invite the reader to compute the two-point functions using the three blocks
picture (\ref{eqIV.II}) and he will discover that he will get precisely the results
presented in Fig.\ref{fXXXI}. 
This applies also to the other two-point functions.
Obviously, the correlation functions dependent only on the
densities but not on $\lambda$.

That in the pure phase one has charge segregation is obvious. Not
only the correlation functions do depend on y in agreement with the
definition given in Sec.\ref{secI}, but the configurations described by (\ref{eqIV.II}) give
the simplest example of segregation. The mixed phase will provide another
example.

\section{The mixed phase ($1< q <q_c$)\label{secV}}
\f78
%
In Sec.\ref{secIII} we have seen that in the mixed phase, for any $\rho$ and $\lambda$,
the current is given by the expression (\ref{eqIII.V}). 
We are going to analyse 
this phase in more detail. In Figure \ref{f78} 
we show for $q=1.2$, $\lambda=1$ and 
$\rho=0.2$ (one can check from Fig.\ref{fII} that we are in the mixed phase) 
the correlation functions $c_{0,0}$ and $c_{+,-}$ 
as a function of $y$ for $L=400$. The ``macroscopic''
variable $y$ is defined in Eq.(\ref{eqIV.III}). The correlation functions were obtained  
using Monte-Carlo simulations. As will be explained in Sec.\ref{secX} 
one can't use
the grand canonical ensemble to compute correlation functions for large distances
(this is not the case for the current) in the mixed or pure phases but only in
the disordered phase.

It is interesting to compare the correlation functions in the mixed
phase and in the pure phase (see Fig.\ref{fXXXI}). 
They look similar. The $c_{0,0}$ function
has the same shape in both cases although the values of $c_{0,0}$ 
are different. For
example the widths of the plateau have different values. 
The $c_{+,-}$ functions
have a slightly different shape
(compare the behaviors for $y<0.4$ and $y>0.7$). One could
be tempted to conclude that there are no major differences between the two phases.
This is not the case. As we mentioned, in the pure phase the data presented
in Fig.\ref{fXXXI} 
represent already the thermodynamical limit. The data presented in 
Fig.\ref{f78} although obtained for a large lattice do not represent the
thermodynamical limit. There is a slow change with $L$ of the behavior of the
correlation functions not shown in the figures. We took data for lattice sizes
up to $L=800$, the $L$ dependence is well taken into account by the mean-field calculations
presented in the next section. 
That these mean-field calculations make sense can already be seen in Fig.\ref{f78}
where the solid curves which are on top of the Monte-Carlo data are 
mean-field results. After presenting the mean-field results we are going 
to compare them with other data obtained from Monte-Carlo simulations (see 
Figs.\ref{fInew} and \ref{fIInew}).

At this point the reader should accept the fact that the correlation
functions converge at large $L$ to non-trivial functions of $y$ (this is going
to be explained in the next section) and therefore in the mixed phase,
like in the pure phase we have charge segregation.

As we are going to prove in Sec.\ref{secVI} 
the major difference between the
two phases is that, as opposed to the pure phase, in the mixed phase translational
invariance symmetry is not spontaneously broken.

\section{Inhomogeneous solutions of the mean-field equations. The pure and mixed 
phases.\label{secIMF}}

We have postponed the presentation of mean-field for the following
reason. We had first to present the empirical data for the correlation
functions in order to know how the mean-field calculations have to be
done. Since for a given lattice size $L$, the correlation functions depend
smoothly on the ``macroscopic'' variable $y$ ($0<y<1$), this suggests to work in
the
continuum taking $y$ as variable and $L$ as a parameter in the mean-field
equations. The previous mean-field calculations in similar problems (see
for example Refs. \cite{citF} and \cite{citG}) were done in a different way. The mean-field 
equations for the stationary state are a couple of non-linear differential
equations and it is not clear a priori how to solve them. The inspiration
came from getting, numerically, the solutions on small lattices. The
results of these investigations as well as the mean-field equations in the
continuum are presented in \ref{appMF}. About the relevance of mean-field
to our problem, we are going to comment later.

In \ref{appMF} it is shown that the for given values of $q$, $\lambda$ and 
densities in the pure or mixed phases the mean-field equations on the 
lattice have besides an homogeneous solution (which was used in Eq.(\ref{eqIII.II})) 
inhomogeneous ones. These solutions are ``pinned'' in space even for a 
finite lattice. Otherwise speaking, in mean-field translational invariance 
is broken in both the pure and mixed phases. These conclusions are based 
on numerical results. One remarkable property of the inhomogeneous solution 
(see Figs.\ref{fCX} and \ref{fCXI}) is that it shows two domains one with charged 
particles only (we called it condensate) and one with charged particles 
and vacancies (we called it fluid). 
In this section we are going to present the stationary solutions of the
mean-field equations (\ref{eqB.III}--\ref{eqB.VI}) which we were able to obtain by making the 
Ansatz of the existence of a condensate and a fluid. A stationary
non-homogeneous solution of the mean-field equations will be 
contain at least one bump. In the present paper we consider the solutions
with one bump only since they are relevant for our problem.
Solutions with more bumps are shortly mentioned in \ref{appMF}
and discussed in detail in \cite{citQ}.
These solutions always breaks translational invariance. We will assume that
mean-field gives the correct description of our physical problem (we will
return to this point at the end of this section) if its solutions are used 
in different ways for the pure and mixed phases.

We will work with  ``macroscopic'' variables like in equation (\ref{eqIV.III}). 
In this case the ring
has a perimeter of length 1. 
For the mixed phase
we will assume that with equal
probability for each point $z$ on the ring ($0<z<1$) one has a bump 
which contains a condensate of length $a$
and a fluid of length $b$ (this implies
that one has translational invariance) and that mean-field describes both of them.
For the pure phase we take the results of mean-field at face value: the 
bump is "pinned" in space and translational invariance is broken.
In both phases 
the values of $a$ and $b$ are $L$-dependent and their values
in the thermodynamical limit will be denoted by $A$ and $B$. Obviously
\begin{eqnarray}
a+b=1\;.
\label{eqV.I}
\end{eqnarray}
It is convenient to choose a coordinate $y$ with the origin at the
beginning of the condensate. Thus, if $0<y<a$ we are in the condensate, and if
$a<y<1$ we are in the fluid. The names condensate and fluid are naturally related
to Bose-Einstein condensation. As will be shown in Sec.\ref{secX} a similar phenomenon 
takes place in our system. 
We assume that
there are no correlations in the condensate and the fluid so that we 
can apply mean-field theory. The present calculation will give not only the 
concentrations of particles but also the values of the current
and the length $b$ of the fluid as a function of $L$. We start with the fluid and
assume that on the segment of length $b$ the concentrations of 
positive and negative particles, $\rho_+^{\rm fl}(L)$ and $\rho_-^{\rm fl}(L)$, 
and implicitly of vacancies,
$\rho_0^{\rm fl}(L)$, are independent of $y$ but $L$ dependent. One has:
\begin{eqnarray}
\rho_+^{\rm fl}(L)+\rho_-^{\rm fl}(L)+\rho_0^{\rm fl}(L)=1
\label{eqV.II}
\end{eqnarray}
In the condensate, one has no vacancies, the concentrations of positive and
negative particles ($\rho_+^{\rm co}(y,L)$ and $\rho_-^{\rm co}(y,L)$) are 
$y$ and $L$ dependent and satisfy
the relation:
\begin{eqnarray}
\rho_+^{\rm co}(y,L)+\rho_-^{\rm co}(y,L)=1
\label{eqV.III}
\end{eqnarray}

We will consider three cases. We take the
same density of positive and negative particles 
\begin{eqnarray}
\rho=\frac{1-v}{2}
\label{eqV.IV}
\end{eqnarray}
and $q$ greater than $1$, next $q=1$ and $q<1$. 
Various predictions obtained from mean-field will be compared with the
results obtained from Monte-Carlo simulations.

\subsection{$q>1$ (mixed phase)\label{secIMFa}}

\subsubsection{Profiles for finite $L$}

\fC

In the fluid we have
\begin{eqnarray}
\rho_+^{\rm fl}=\rho_-^{\rm fl}=\frac12 (1-\rho_0^{\rm fl})
\label{eqV.V}
\end{eqnarray}
and 
\begin{eqnarray}
b \rho_0^{\rm fl}=v
\label{eqV.VI}
\end{eqnarray}
The current is
\begin{eqnarray}
J=\lambda \rho_+^{\rm fl}\rho_0^{\rm fl}+(q-1) \rho_+^{\rm fl}\rho_-^{\rm fl}
\label{eqV.VII}
\end{eqnarray}
For later convenience, it is useful to define
\begin{eqnarray}
\alpha=\sqrt{\frac{4J}{q-1}-1}
\label{eqV.VIII}
\end{eqnarray}
It is clear that if the current has the expression (\ref{eqIII.V}), this implies $\alpha=0$ and that
non-zero values for $\alpha$ 
give finite-size corrections. Using Eqs.(\ref{eqV.V})-(\ref{eqV.VIII})
we get:
\begin{eqnarray}
\alpha^2=\frac{v}{b(q-1)} \left( 2(\lambda+1-q)+\frac{v}{b}(q-1-2\lambda)\right)
\label{eqV.IX}
\end{eqnarray}
We define
\begin{eqnarray}
B=\frac{v(1-q+2\lambda)}{2(1-q+\lambda)}\,.
\label{eqV.X}
\end{eqnarray}
As will be seen, 
$B$ will represent the large $L$ limit of the length $b$ of the fluid.
Denoting
\begin{eqnarray}
\gamma=\frac{v}{B}-\frac{v}{b}
\label{eqV.XI}
\end{eqnarray}
and using Eq.(\ref{eqV.IX}) we obtain
\begin{eqnarray}
\alpha^2=\frac{2(1+\lambda-q)}{q-1}\gamma\left(1-\frac{1-q+2\lambda}{2(1-q+\lambda)}\gamma\right)
\label{eqV.XII}\;.
\end{eqnarray}
Eq.(\ref{eqV.XII}) connects the value of the current and the width of the fluid. We are
going to obtain a second equation relating the same quantities examining the
condensate. We write the expression of the current in the condensate in the
mean-field approximation in a symmetric way
\begin{eqnarray}
J=\frac{q}{2}\left(\rho_+^{\rm co}(y-\frac{1}{L})\rho_-^{\rm co}(y)+\rho_+^{\rm co}(y)\rho_-^{\rm co}(y+\frac{1}{L})\right)\nonumber\\
\qquad
-\frac{1}{2}\left(\rho_-^{\rm co}(y-\frac{1}{L})\rho_+^{\rm co}(y)+\rho_-^{\rm co}(y)\rho_+^{\rm co}(y+\frac{1}{L})\right)
\label{eqV.XIII}
\end{eqnarray}
and define
\begin{eqnarray}
\rho_\pm^{\rm co}=\frac{1}{2}\pm w\,.
\label{eqV.XIV}
\end{eqnarray}
Eq.(\ref{eqV.III}) is then satisfied for any $w$. 
Using Eqs.(\ref{eqV.XIII}) and (\ref{eqV.XIV}) and going 
to the continuum limit we find the following expression for the current:
\begin{eqnarray}
J=(q-1)(\frac{1}{4}-w^2)-\frac{q+1}{2L}\frac{{\rm d} w}{{\rm d}y}
\label{eqV.XV}
\end{eqnarray}
(an analogous expression for the current appeared for the first time in
Ref. \cite{Krug1})
The same relation could have been obtained directly from Eq.(\ref{eqB.III})
We keep in mind that charge conservation gives
\begin{eqnarray}
\int_0^a \rho_\pm^{\rm co}\,\d y=\frac{a}{2}\,.
\label{eqV.XVI}
\end{eqnarray}
We will now impose boundary conditions for the differential equation (\ref{eqV.XV}). We
will require that the concentration of positive particles is continuous at the 
end of the condensate ($y=a$) and that of the negative particles at the beginning 
of the condensate ($y=0$):
\begin{eqnarray}
\rho_+^{\rm co}(a)=\rho_+^{\rm fl} \nonumber\\
\rho_-^{\rm co}(0)=\rho_-^{\rm fl}
\label{eqV.XVII}
\end{eqnarray}
Notice that in this way, the density of positive particles $\rho_+(y)$ is
continuous at the point $a$ but discontinuous in the origin
where one has a shock even at finite $L$. Conversely, the
density of negative particles $\rho_-(y)$ is continuous for $y=0$ but not for
$y=a$.
Using Eq.(\ref{eqV.XV}) and (\ref{eqV.XVII}) 
we get the profile of the concentrations in the
condensate:
\begin{eqnarray}
w=\frac{\alpha}{2}\tan\left(\beta(\frac{a}{2}-y)\right)
\label{eqV.XVIII}
\end{eqnarray}
where $\alpha$ is given by Eq.(\ref{eqV.VIII}), and $\beta$ is related to $\alpha$:
\begin{eqnarray}
\beta=\frac{q-1}{q+1}L\alpha
\label{eqV.XIX}
\end{eqnarray}
The boundary conditions give:
\begin{eqnarray}
\beta\tan(\frac{\beta a}{2})=\frac{q-1}{q+1} \frac{v}{1-a}L
\label{eqV.XX}
\end{eqnarray}
The equation (\ref{eqV.XX}) relates again the current to the width of the fluid. In 
this way, from the Eqs.(\ref{eqV.XII}) and (\ref{eqV.XX}) 
we can derive both the current $J$ and $b$
as functions of $v$, $\lambda$, $q$ and $L$. 
Once they are known, using (\ref{eqV.VI}) and (\ref{eqV.XVIII})
one can derive the concentrations in the condensate and the fluid. 
Knowing the
concentrations, one can compute any correlation function like 
$c_{0,0}$, $c_{+,-}$,
etc.

Notice that for $y=0$ the density $\rho_+$ is discontinuous even for finite $L$.
Therefore the bump ends with a shock on the left.
As we are going to show shortly the bump will end with a shock also on the right
in the large $L$ limit.

In order to solve the transcendental equations (\ref{eqV.XII}) 
and (\ref{eqV.XX}), we will
do a large $L$ expansion. It is convenient to define a new variable $u$ instead of
$\beta$:
\begin{eqnarray}
\beta=\frac{\pi-2u}{a}
\label{eqV.XXI}
\end{eqnarray}
and instead of Eq.(\ref{eqV.XX}) we obtain:
\begin{eqnarray}
\frac{\tan u}{1-\frac{2u}{\pi}}=\delta(1-a)v^{-1}
\label{eqV.XXII}
\end{eqnarray}
where
\begin{eqnarray}
\delta=\frac{\pi(q+1)}{L a (q-1)}\,.
\label{eqV.XXIII}
\end{eqnarray}
Keeping only the term of order $u$ in the left-hand side of Eq.(\ref{eqV.XXII})
and using
Eqs. (\ref{eqV.XIX}) and (\ref{eqV.XXI}) we obtain:
\begin{eqnarray}
\alpha^2=\delta^2(1-\frac{4(1-a)}{v\pi}\delta)
\label{eqV.XXIV}
\end{eqnarray}
We now use Eq.(\ref{eqV.XII}) to obtain:
\begin{eqnarray}
\gamma=\frac{\delta^2(q-1)}{2(1+\lambda-q)}
\label{eqV.XXV}
\end{eqnarray}
Collecting all the results together, we obtain finally:
\begin{eqnarray}
\frac{4J}{q-1}=1+\left(\frac{(q+1) \pi}{(q-1) A L}\right)^2\left(1-\frac{4(q+1)(1-A)}{L(q-1)v A}\right)
\label{eqV.XXVI}
\end{eqnarray}
and
\begin{eqnarray}
\frac{b-B}{v}=\left(\frac{B}{v}\right)^2 \frac{(q+1)^2 \pi^2}{2(1+\lambda-q)(q-1)A^2 L^2}
\label{eqV.XXVII}
\end{eqnarray}
where
\begin{eqnarray}
A=1-B\,.
\label{eqV.XXVIII}
\end{eqnarray}
$B$ is the length of the fluid in the thermodynamic limit and $A$ is the length of
the condensate. From Eq.(\ref{eqV.XXVI}) we learn that in the large $L$ 
limit the current
is indeed given by Eq.(\ref{eqIII.V}) and 
that the finite-size correction terms are not
only $q$ dependent but (through $A$) are dependent also on the density of vacancies $v$
and on $\lambda$. Notice that the finite-size corrections to the current although
quadratic in $1/L$ (the relevant parameter is $\delta$ given by Eq.(\ref{eqV.XXIII})) can be
large even for large lattices, say $L=200$. We remind the reader that in Sec.\ref{secIII} we
have shown that in the grand canonical ensemble the finite-size
corrections are linear in $1/L$. A different finite-size behavior for the two
ensembles was already noticed for the disordered phase (see Eqs.(\ref{eqIII.IX}-10)).


%
\fInew		      
\fIInew		      

\subsubsection{Comparison of the mean-field predictions with results obtained by other methods.
  Finite-size scaling.}

We are now in the position to compare the result of the model with the Monte-
Carlo data. We first start with the current (see Fig.\ref{fIVa}). We notice that for
large values of $L$ ($L>150$) the mean-field values obtained using the method of
Appendix B which are already almost identical with those obtained in the 
continuum approximation given above are close to the Monte-Carlo data. One can
be more precise: Using Eq.(\ref{eqV.XXVI}) with $q=1.2\, \lambda=1\,\rho=0.2$ one gets 
$J=0.0534$ for $L=200$ and $J=0.0522$ for 
$L=400$ to be compared with the values $J=0.059$ respectively $J=0.053$. This
observation is interesting since one wouldn't expect a mean-field calculation
to control so well finite-size corrections. For smaller values of $L$ (see again
Fig.\ref{fIVa}), the Monte-Carlo data (confirmed by the calculations done using the
algebraic approach) as well as the mean-field values show a strange behavior of 
the current in the region $50<L<70$. The explanation of
this phenomenon is simple. For given values of the parameters $q$, $\lambda$ and $\rho$
for which in the thermodynamical limit one is in the mixed phase, the bump 
picture appears only if the size $L$ of the system is larger than a certain 
minimal value.
We would like to add that we have not checked for which values of $L$ the continuum version
of mean-field coincides with the discrete version presented in Appendix B.

The phase diagram is at this point known. The mixed phase, for a given values
of $\lambda$ and $\rho$, is between $1<q<q_c$, where $q_c$ is 
\begin{eqnarray}
q_c=1+\frac{2\lambda(1-v)}{2-v}\,.
\label{eqV.XXIX}
\end{eqnarray}
This value is obtained from the condition
\begin{eqnarray}
B=1
\label{eqV.XXX}
\end{eqnarray}
which says that one has no condensate. 

\tI
One can ask to which extent the mean-field prediction (\ref{eqV.XXIX}) is correct.
In Table~\ref{tI} we present the estimates of the critical points $q_c^{gc}$ obtained
from the ``discontinuities'' of the derivatives of the current and of the
two-point functions $c_{-,+}(1)$ as a function of $q$ using the grand canonical
ensemble (see Sec.\ref{secIII}) together with the values of $q_c$ obtained using formula 
(\ref{eqV.XXIX}). The estimates have errors which are hard to evaluate, therefore an
optimistic point of view would suggest that (\ref{eqV.XXIX}) is exact. A pessimist
would say that mean-field is a good approximation in the vicinity of
the phase transition only for low values of $\lambda$ and densities. An
argument in favor of the optimistic point of view will be given
below.

A well known question in equilibrium statistical physics is the problem of
finite-size scaling corrections in the determination of critical points \cite{citL}. 
One can ask a similar question here. For a finite value of $L$ (at fixed 
$\lambda$ and $\rho$) an estimate of $q_c$ can be obtained for example using the 
fact that in the thermodynamical limit the derivative of the current has a 
a jump at $q_c$ which at finite values of $L$ gives an obvious way to get an 
estimate for $q_c$. We would like to know how the estimate approaches
$q_c$. In order to get an insight to this problem, one can assume that 
the current changes when the condensate appears. Therefore let
us go back to Eq.(\ref{eqV.XXVII}) which gives
the length of the condensate for finite $L$. One sees that $b>B$. 
In particular, for 
a given value of $L$, there is a value of $q$ that we denote by $\tilde{q}_c(L)$ for which
there is no condensate, i.e.
\begin{eqnarray}
b(L)=1\,.
\label{eqV.XXXI}
\end{eqnarray}
Denoting
\begin{eqnarray}
\tilde{q}_c(L)=q_c-\epsilon
\label{eqV.XXXII}
\end{eqnarray}
and using Eqs.(\ref{eqV.XXVII}) and (\ref{eqV.XXXI}) we obtain:
\begin{eqnarray}
\epsilon^3 L^2= \frac{8 \lambda v \pi^2}{(2-v)^6(1-v)}(2-v+\lambda(1-v))^2
\label{eqV.XXXIII}
\end{eqnarray}
From Eq.(\ref{eqV.XXXIII}) we see that $\epsilon>0$, which means that for a given value of
$q<q_c$ up to a certain length $L$, one has no bump, in agreement with the 
interpretation given above to the behavior of the current as a function of $L$
in Fig.\ref{fIVa}. On the other hand Eq.(\ref{eqV.XXXIII}) 
gives obviously a finite-size scaling
critical exponent $2/3$.
Unfortunately, this derivation is only heuristic. Here are the reasons. A 
careful examination of the Eqs.(\ref{eqV.XVIII}--\ref{eqV.XX}) shows that the value $b(L)=1$ is 
never obtained although the large $L$ approximation (\ref{eqV.XXVII}) gives it. Mean-field 
theory suggests that the condensation appears directly with a non-zero value 
of the length $a$ and implicitly a value of $b<1$. Using this value of $b$ 
(which is obtained from a transcendental equation) one can estimate
numerically the critical exponent.
It is also $2/3$ but the right hand side of Eq.(\ref{eqV.XXXIII}) is different. 

A careful examination of the Monte-Carlo data shows a 
similar phenomenon although the value of $a$ at which the condensate appears is slightly different than the 
one predicted by mean-field.

Instead of working at fixed $v$ and $\lambda$, one can take fixed $q$, $\lambda$ (in 
the mixed phase) and $L$ and look for at which density $\bar{v}_c(L)$ starts the 
condensation ($b(L)=1$). We use again Eq.(\ref{eqV.XXVII}) and get 
\begin{eqnarray}
L^2\left(1-\frac{\bar{v}_c(L)}{v_c}\right)^3
=\pi^2 \frac{(q+1)^2}{(q-1)(1-q+2\lambda )}
\label{eqV.A}
\end{eqnarray}
where $v_c$ is obtained from Eq.(\ref{eqV.X}) with $B=1$:
\begin{eqnarray}
v_c=\frac{2(1-q+\lambda)}{1-q+2\lambda}
\label{eqV.B}
\end{eqnarray}
From Eq.(\ref{eqV.A}) we learn that for finite $L$ the condensation occurs at a lower 
density of vacancies (larger density of particles) than at the critical 
point. This behavior can also be seen in Fig.\ref{fC} where the $\rho$ dependence 
of the current at fixed $q=1.2$ and $\lambda=1$ is shown for various lattice sizes.
For large values of $\rho$ one is in the mixed phase ($J =(q-1)/4$). If we 
decrease the value of $\rho$ the current increases (see Eq.(\ref{eqV.XXVI})) up to a 
point where the condensate disappears. For small values of the 
density the mean-field approximation does not apply (one is in the 
disordered phase). If one increases the size of the lattice, the minimal 
value of $\rho$ where mean-field still applies gets smaller and the current gets
closer to its asymptotic value $(q-1)/4$. One way to estimate $v_c$ is to 
take the values of the density $\bar{\rho}_c$ where the current has a 
maximum. One can then estimate the exponent alpha from the expression
\begin{eqnarray}
1-\frac{\bar{v}_c(L)}{v_c}={\rm const.} L^{-\alpha}
\label{eqV.C}
\end{eqnarray}
Using data taken at $L=100,\,200,\,400,\,800$ and $1600$ one gets $\alpha=0.5\pm0.2$
which makes the exponent $\alpha=2/3$ given
by Eq.(\ref{eqV.A}) plausible.
Moreover, this also implies that the mean-field value for $v_c$ given by
Eq.(\ref{eqV.B}) is probably correct.  
The estimate of $\alpha$ has large errors since it is difficult to determine precisely
the position of the maximum of the current.

We are going to verify the mean-field predictions in the mixed phase away from $q_c$. 
We take again $q=1.2$, $\lambda=1$,
$\rho=0.2$ and $L=400$. 
Solving Eqs.(\ref{eqV.XII}) and  (\ref{eqV.XX}) numerically 
and using Eqs.(\ref{eqV.XVIII}) and (\ref{eqV.XIX}) we get:
\begin{eqnarray}
\fl
J=0.05257 ,\quad   a=0.3200 , \quad  w= 0.1134\tan\left(8.244 (\frac{a}{2}-y)\right) 
\label{eqV.XXXIV}
\end{eqnarray}
We have done Monte-Carlo simulations taking only configurations in which $a$ had 
approximately the value given in (\ref{eqV.XXXIV}). 
For these configurations we have 
determined the concentrations of positive and negative particles. They are shown
in Fig.\ref{fInew} together with the 
prediction obtained from 
Eq.(\ref{eqV.XXXIV}). 
The agreement couldn't be better.  One can use (\ref{eqV.XXXIV}) in order to compute the correlation functions 
$c_{0,0}$ and $c_{+,-}$
and compare them with the Monte-Carlo data. This comparison can be seen in Fig.\ref{f78}
and was repeated for all correlation functions. The agreement is as good as
for $c_{0,+}$ and $c_{-,+}$.

\subsubsection{Profiles for large $L$}

We would like to discuss now the composition of the bump in the thermodynamical
limit. We denote the concentrations by $R$ instead of $\rho$. From Eqs.(\ref{eqV.XIV}), (\ref{eqV.XXIV})
and (\ref{eqV.XVII}), we get
\begin{eqnarray}
R_{\pm}^{\rm co}(y)=\frac12,\qquad R_0^{\rm fl}=\frac{v}{B},\qquad
R_{\pm}^{\rm fl}=\frac{1-R_0^{\rm fl}}{2}
\label{eqV.XXXV}
\end{eqnarray}
This result is mostly interesting. One can see the closed system as two open systems, one
containing only positive and negative particles (the condensate) with equal 
concentrations and a second one containing also vacancies (the fluid). From our
knowledge of the two-state model \cite{citR} with the bulk rates 
$g_{+,-}=q,\, g_{-,+}=1$
we know that
the current has the expression (\ref{eqIII.V}) from mean-field (it is just a
repetition of the calculation done above where Eq.(\ref{eqV.XVII}) is replaced by an 
equation which takes into account the boundary rates). This is the so-called 
maximum-current phase \cite{citO,citS}. 
In the same maximum-current phase but only for the
case of totally asymmetric diffusion it was shown that the correlation functions
have an algebraic decay. It is plausible to assume that the same applies
for the partially asymmetric case which is ours. This would also suggest that
the fluid phase is also algebraic. This assumption can be checked independently
since the algebraic tools for this problem are already known \cite{citT}. Unfortunately
we were not able to check if indeed the correlation functions $c_{0,0}$, $c_{+,-}$ etc.
in the present model have an algebraic decay. This would imply to work not with the macroscopic
variable $y$ but with $R$ (see Eq.(\ref{eqIV.III})) 
and to take lattice sizes beyond our means.

The reader might have noticed that all the expansions used above diverge for
$q=1$ (see Eq.(\ref{eqV.XXIII})). We proceed now to discuss this case.

\subsection{$q=1$\label{secIMFb}}

A repetition of the calculation done for the case $q=1$ gives the following 
results. The current in the fluid phase has the expression:
\begin{eqnarray}
J=\frac{\lambda}{2}(1-\frac{v}{b})\frac{v}{b}
\label{eqV.XXXVI}
\end{eqnarray}
In the condensate, the current is 
\begin{eqnarray}
J=-\frac{1}{L} \frac{{\rm d}w}{{\rm d}y}
\label{eqV.XXXVII}
\end{eqnarray}
which can be obtained from Eq.(\ref{eqV.XV}) for $q=1$. 
Taking into account the boundary
conditions, we obtain:
\begin{eqnarray}
w=\frac{v}{ab}(\frac{a}{2}-y)
\label{eqV.XXXVIII}
\end{eqnarray}
and 
\begin{eqnarray}
J=\frac{v}{a b L}
\label{eqV.XXXIX}
\end{eqnarray}
Equating the two expressions for the current (\ref{eqV.XXXVI}) and (\ref{eqV.XXXIX}), 
we find $b$ as a
function of $L$ and consequently, $J$ as a function of $L$. In order to show how good
our results are, one can compare for $\lambda=1$, $\rho=0.2$ and $L=100$ (which is a
small lattice) the value of the current obtained using (\ref{eqV.XXXVI})-(\ref{eqV.XXXIX}) which is
$J=0.02588$ with the Monte-Carlo value $J=0.0256$.

In Fig.\ref{fIInew}, we compare the profiles, now for $q=1$, just as we did
in Fig.\ref{fInew} for $q>1$. Notice the nice straight lines.

For large values of $L$ one obtains the following leading contributions to the
current
\begin{eqnarray}
J=\frac{1}{(1-v) L}-\frac{2(1-2v)}{\lambda(1-v)^3 L^2}
\label{eqV.XXXX}
\end{eqnarray}
and to the length of the fluid
\begin{eqnarray}
b=v\left(1+\frac{2}{\lambda L (1-v)}\right)
\label{eqV.XXXXI}
\end{eqnarray}
which explains Eq.(\ref{eqIII.XII}).

In the thermodynamic limit, keeping the notations used already, we get:
\begin{eqnarray}
B=v,\qquad A=1-v
\label{eqV.XXXXII}\\
R_+^{\rm fl}=0, \qquad R_+^{\rm co}=1-\frac{y}{A}
\label{eqV.XXXXIII}\\
R_-^{\rm fl}=0, \qquad R_+^{\rm co}=\frac{y}{A}
\label{eqV.XXXXIV}
\end{eqnarray}
and of course $J=0$.

It should be interesting to study in detail the properties of the fluid and
the condensate at $q=1$ looking not at macroscopic distances ($y$) but at microscopic
ones ($R$).

\subsection{$q< 1$ (pure phase)\label{secIMFc}}

One can repeat the calculations done for the mixed phase for the pure phase.
We give directly the results. Instead of Eq.(\ref{eqV.XVIII}) one has:
\begin{eqnarray}
w(y)=\frac12\left({\frac{4J}{1-q}+1}\right)^{1/2} \tanh\left(\frac{q-1}{q+1} L \left({\frac{4J}{1-q}+1}\right)^{1/2}  (\frac{a}{2}-y) \right)
\label{eqV.D}
\end{eqnarray}
which gives the profiles of the concentrations in the condensate, the 
boundary condition (\ref{eqV.XX}) been replaced by 
\begin{eqnarray}
w(0)=\rho^{\rm fl}_0=v/b
\label{eqV.E}
\end{eqnarray}
This allows to obtain for large values of $L$ the expression of the current:
\begin{eqnarray}
J=\frac{\lambda(1-q)}{\lambda+1-q}\exp \left( \frac{q-1}{q+1} L (1-v)\right)
\label{eqV.F}
\end{eqnarray}
and of the density of vacancies in the fluid:
\begin{eqnarray}
\rho^{\rm fl}_0=1-2J/\lambda
\label{eqV.G}
\end{eqnarray}
the last expression shows, as expected, that in the large $L$ limit, the 
densities of vacancies in the fluid tends exponentially to $1$ which is 
precisely what one expects in the fluid phase.
The expression of the current given by Eq.(\ref{eqV.F}) is compatible with the 
fits done to the Monte-Carlo data (see for example (\ref{eqIII.XI})).
An exponential fall-off of the current was also observed in a different
model \cite{Ramaswamy}.

\section{The spontaneous breaking of translational invariance.
Flip times in the pure and mixed phase \label{secVI}}

\fXVIII
\fXVII
\fXIX
In the last sections we have described the pure and the mixed phases. Both
phases show long range correlations which are different from a quantitative
point of view (exponential versus algebraic finite-size convergence 
etc.), but up to this point we haven't proven that they are also 
qualitatively different. Moreover, mean-field calculations (see \ref{appMF} and Sec.\ref{secIMF})
show no qualitative differences between the two phases: one obtains pinned 
configurations in both cases. We are going to show now that half of these mean-field 
predictions are wrong and that in the pure phase translational invariance 
is broken and that this is not the case for the mixed phase.

It is useful to consider the order parameter $M$ defined as follows:
\begin{eqnarray}
M=\frac1L \sum_{k=1}^{L} \exp(\frac{2 \pi\i}{L}k) \delta(\beta_k)
\label{eqVI.I}
\end{eqnarray}
One notices that the order parameter is sensitive only to the vacancies 
and to the macroscopic properties of the system (it sees only the lowest 
frequency). For a given length $L$ of the system we have measured using 
Monte-Carlo simulations the probability $P(|M|)$ to find a certain value 
of the modulus of $M$. Next we have considered the \FEF 
\begin{eqnarray}
\label{eqVI.II}
f(|M|)=-\frac1L \log P(|M|)
\end{eqnarray}
In Ref.\cite{citU} we have proven, for another order parameter and another model, 
that this function converges in the thermodynamical limit. In Figs.\ref{fXVIII},
\ref{fXVII} and \ref{fXIX} this function is shown for the pure, mixed and disordered phases. 

These figures confirm know results. Indeed if one has a block of vacancies of 
length $b$ (again macroscopic length!) a trivial calculation shows that 
$f(|M|)$ should have a minimum at the value
\begin{eqnarray}
|M|_{\rm min}=\frac{v}{\pi b} \sin(\pi b) \; .
\label{eqVI.III}
\end{eqnarray}
Here $v$ is as usual the density of vacancies. In the pure phase ($b=v$) this 
gives the value $|M|_{\rm min} =0.3027$,  for the mixed phase one can use the value $b= 0.675$ 
obtained using Eq.(\ref{eqV.X}) which gives $|M|_{\rm min}=0.238$ and, finally, for 
the disordered phase one gets $|M|_{\rm min}=0$ since here $b=1$ in agreement with 
the values seen in the figures.

\fXIV
\fXV
We would like at this point to make, what is in our view, an important 
observation. In Ref.\cite{citU} it was advocated that the \FEF\ defined 
in Eq.(\ref{eqVI.II}) is a useful concept not only in equilibrium statistical
physics but also in the present context. As one can see from Figs.\ref{fXVII} and \ref{fXIX}
this is certainly the case for the mixed and disordered phases since this 
function nicely converges for large $L$. This is not the case for the pure phase.
Notice that in this case the function which converges is rather $f(|M|)/L$.

A similar phenomenon occurs for another concept ``exported''
from equilibrium statistical physics to non-equilibrium, namely the
position of the roots of the partition function seen as a function of
fugacity \cite{LeeYang}. As shown in Ref. \cite{Peter}, they converge to a parabola in
the disordered phase, to an ellipse in the mixed phase,
but shrink for large $L$ for the pure phase.
These
observations might be relevant for other models or for the same model in two
dimensions since it might give a way to distinguish between the pure and mixed
phases. 

In order to obtain new information one can look not to $|M|$ but to $M$ 
itself and, instead of considering the stationary state only,
examine the time dependent behavior of the system. 
Using Monte-Carlo simulations we have looked for a 
given size $L$ of the system, at the average flip time $T$ from a configuration
which is in the domain  ${\rm Re}(M)>0,\; -0.1<{\rm Im}(M)<0.1$ to a 
configuration ${\rm Re}(M)<0$, 
$-0.1<{\rm Im}(M)<0.1$. 
Here ${\rm Re}(M)$ and ${\rm Im}(M)$ represent the real respectively the imaginary
parts of $M$. These average flip 
times are shown as functions of the size of the system $L$ in Fig.\ref{fXIV} (for 
the pure phase) and in Fig.\ref{fXV} (for the mixed phase). 
For the pure phase we limited ourselves to $L=140$ since for larger values of
$L$ the flip times are to long.
Now we notice a major
difference between the
two phases: whereas in the pure phase the flip time increases exponentially 
with $L$ (exponentially increasing barrier), in the mixed phase the flip time 
increases algebraically. 
Actually the data are compatible with a linear dependence. 
This result implies that in the pure phase but 
not in the mixed phase, translational invariance is spontaneously broken. 
This settles the problem about the difference between the pure and the mixed
phase. 

If one uses as a definition of charge segregation the existence of correlation
functions which are dependent of the macroscopical variables ($y$ in our
notations for the two-point functions), we have to distinguish between the two
possibilities put in evidence in this section. One in which one has the 
breaking of translational invariance, which we propose to call ``charge 
segregation of type B'' and one in which translational invariance is not broken
which we propose to call ``charge segregation of type UB''. 

We would like to mention that another approach to average flip
times in order to put in evidence charge segregation can lead to ambiguous
results. A different model was studied in Ref.\cite{citE}. The authors looked,
(this would correspond to our model), at the average flip time $T$ from a
configuration with $b_1 |M|_{\rm min}$ to another with $b_2 |M|_{\rm min}$. Here $b_1$ and $b_2$ are $L$
independent constants. They found that $T$ increases exponentially with $L$.
From the shape of the \FEF\ shown in Figs.\ref{fXVIII} and \ref{fXVII}
the same should be true also in our case for both the pure and mixed 
phases. We could distinguish between the pure and mixed phases only by
considering the average flip times in the complex $M$ plane.

We can now continue the analysis of the model and consider the 
disordered phase.

\section{The disordered phase.\label{secVII}}
\fXbbb
\fXbb
\fXXII

For fixed $\lambda$ and densities and $q>q_c$ where $q_c$ is given by Eq.(\ref{eqV.XXIX}),
one finds the disordered phase. In this phase the correlation length is
finite. What is the information we have already about this phase?

a) For large values of $q$, the current is known (see the large $q$ expansion
presented in Sec.2) and is given in the case of equal densities of the
charged particles by Eqs.(\ref{eqII.XXXXVIII}), (\ref{eqII.XXXXXV}) and (\ref{eqII.XXXXXVIII}).

b) For $\lambda$ and $q$ related by the relations (\ref{eqII.XII}) (those are curves in
the $q$--$\lambda$ plane in the domain where one is in the disordered phase), the
quadratic algebra (\ref{eqII.III}) has finite-dimensional representations. If the
representation is finite-dimensional, it follows that the
correlation length is finite. The calculation of various correlation
functions is simple but nevertheless tedious. We have not done any
calculation of this kind because we didn't expect any surprises. 

Since we are interested to understand the nature of the phase transition
at $q_c$, we have taken $\lambda=1$, $\rho=0.2$ (for this case we know, using
Eq.(\ref{eqV.XXIX}), that $q_c=1.57$) and computed the correlation function 
$c_{+,-}$ as a function of the microscopic distance $k$ for three values of $q$. We
have taken $L=100$ and used the grand canonical ensemble. The results are
shown in Fig.\ref{fXbbb} in a double-logarithmic scale. One sees that in the
disordered phase ($q=1.7$ and $1.8$) one has an exponential fall-off but that
at $q_c$ one has an algebraic one. A fit to the data gives:
\begin{eqnarray}
c_{+,-}(k)=(0.04\pm 0.01)k^{-1.2\pm 0.3}
\label{eqVII.I}
\end{eqnarray}
Next, we were interested in
the $q$ dependence of the correlation length. For this purpose, for a
given value of $q$, we have, using the grand canonical ensemble, taken
lattices of different lengths $L$, and determined the correlation length
$\xi$ fitting $\xi$ of equation
\begin{eqnarray}
c_{+,-}(k)=A\,k^{-1}\,\exp(-k/\xi) 
\label{eqVII.II}
\end{eqnarray}
The factor $k^{-1}$ was chosen in order to improve the fits to the data.
Then we have extrapolated the values of $\xi$ to obtain the large $L$ limit
of $\xi$. These values are shown in Fig.\ref{fXbb}. As expected the inverse
correlation length vanishes at the critical point. As seen from the
figure, the data are compatible with the fit:
\begin{eqnarray}
\xi^{-1}=0.64\,(q-q_c)^{0.6}
\label{eqVII.III}
\end{eqnarray}
At $q_c$, the correlation length has the following large $L$ behavior:
\begin{eqnarray}
\xi= 0.0065 L + 3.98     
\label{eqVII.IV}
\end{eqnarray}
Taking into account that one obtains an excellent fit to the data, the
readers who like conformal invariance should give a second thought to this
result.

It is obvious that more work is needed in order to fix the values of the
two exponents in Eqs.(\ref{eqVII.I}) and (\ref{eqVII.III}), check for universality, etc.
Our
attention however was taken by a problem which we have considered more
interesting, namely to understand which symmetry is spontaneously broken
at the phase transition. We remind the reader, that the symmetries of the
system are translational invariance (for which we have shown in Sec.\ref{secVI} that it
is not broken in both the disordered and mixed phase), $CP$ and $U(1)\times U(1)$ corresponding to the conservation of the
numbers of positive and negative particles. The answer to this question is
given in the next figure.

In Fig.\ref{fXXII} we show for $\lambda=1$ and $q=1.2$ (which for $\rho=0.2$ corresponds
to the mixed phase but becomes $q_c$ if $\rho=0.055$), the fugacity($z$)
dependence on the density($\rho$) for three values of $L$. One notices that
the $L$ dependence of the curves suggest that in the thermodynamical limit, $\rho$
increases linearly with $z$ up to the point where $\rho$ reaches the value
0.055 (the linear increase is an empirical observation). At the
corresponding value of $z$, one gets a vertical and $z$ does not
fix anymore the density. Such a behavior of the density is known in
Bose-Einstein condensation \cite{citV} where (there one has a $U(1)$ symmetry)
the conservation of the number of particles is broken. In the next
section we are going to discuss this topic in detail.

\section{Spatial condensation\label{secX}\label{secVIII}}

\fXVI
\fXXX
\fCI

As is well known, Bose-Einstein condensation is a quantum mechanical
phenomenon in which identical particles condensate in a state of zero momentum.
This phenomenon takes place in three or more dimensions. In a simple
but fascinating one-dimensional stochastic model with a ``defect'' (reviewed in Appendix C), M. Evans \cite{citW}
has shown that, choosing the right variables, the probability
distribution for the stationary state can be written in terms of
occupation numbers with a ``dispersion'' relation (they are no momenta
involved) which give the equivalent of a Bose-Einstein condensation (there 
is a zero bosonic mode). In this model the condensate corresponds to an
empty domain (no particles, just vacancies). 

We will pursue the investigation of our model in order to see if indeed
more properties are common to the usual Bose-Einstein condensation and to
the transition at $q_c$. The compressibility $\kappa$ is defined by the relation:
\begin{eqnarray}
\kappa=\zeta^2 L
\label{eqVIII.I}
\end{eqnarray}
where
\begin{eqnarray}
\zeta=\frac{<\!(\Delta\rho)^2\!>^{1/2}}{\rho}
\label{eqVIII.II}
\end{eqnarray}
Using Eq.(\ref{eqII.XXXVI}) we have measured $\zeta$ as a function of $L$ for $\lambda=1$,
$\rho=0.2$ and several values of $q$. The data are shown in Fig.\ref{fXVI}.
We notice that for $q=0.9$ (in the pure phase) and $q=1.4$ (in the mixed
phase), $\zeta$ tends to a constant which gives a divergent compressibility.
This is not the case in the disordered phase ($q=2.2$). At the critical
point ($q_c=1.6$) $\zeta$ does not decrease like $L^{-1/2}$ but the lattice sizes
are too small in order to decide its ultimate behavior.

Another quantity of interest is the ``pressure'' $P$, which can be defined as for equilibrium statistical physics:
\begin{eqnarray}
P=\frac1L \log Z^{\mu,\mu}_L
\label{eqVIII.III}
\end{eqnarray}
and can be determined using Eq.(\ref{eqII.XXXXI})
\begin{eqnarray}
J=\lambda z \frac{  Z^{\mu,\mu}_{L-1}}{  Z^{\mu,\mu}_L}
=\lambda z (1-\frac{\d}{\d L}  \log Z^{\mu,\mu}_L)
\label{eqVIII.IV}
\end{eqnarray}
Assuming that $\log Z_L^{\mu,\mu}$ is an extensive quantity one obtains: 
\begin{eqnarray}
P=1-\frac{J}{\lambda z}
\label{eqVIII.V}
\end{eqnarray}
In order to find an ``isotherm'' (there is no temperature in our problem) we
have, for $q=1.2$ and $\lambda=1$, computed the current as a function of $\rho$ (at
each value of $\rho$ we have have computed the current for various lattice
sizes and extrapolated to large $L$). The ``pressure'' was computed using
Eq.(\ref{eqVIII.V}) and is shown in Fig.\ref{fXXX}. There are several interesting things that
we can learn from this figure. First that the ``pressure'' is positive, next
that it decreases when the density gets smaller. There are no theoretical
reasons that we can think of that can justify why the ``pressure'' defined
by Eq.(\ref{eqVIII.V}) should have the same properties as the pressure in equilibrium
systems. Moreover, the ``isotherm'' shown in Fig.\ref{fXXX} looks similar to the
isotherm of the free Bose gas when Bose-Einstein condensation takes place. 

\fXc
\fXXIX
\fXXVIII

Having in view other systems where spatial condensation might take place
we have looked for another method to put it into evidence. We have looked 
at a given lattice size $L$ at the probability to find a local value of the 
density of vacancies $v_l$ which is defined as a local average in 
the macroscopic $y$ coordinate ($v_l=$ number of vacancies in the interval 
$\Delta y$ divided by $\Delta y$). The corresponding \FEF\ $f(v_l)$ 
for various values of $L$ with $q$ in the disordered phase and at the
critical point is shown in 
in Fig.\ref{fCI}. 
In the disordered phase ($q=2.2$) $f(v_l)$ has, as it should, a
minimum at $v=0.6$ which corresponds to the average density. 
For $q=q_c$ the function
$f(v_l)$ is almost flat which would give a
\FEF\  corresponding to a first-order phase transition 
although we know that the transition is second-order. A similar situation
occurs also in the case of Bose-Einstein condensation where some
thermodynamical quantities behave like in a first-order phase transition \cite{citV}.

At this point one can ask oneself to which extend the grand canonical
ensemble is useful at all in the mixed phase (we have used it
consistently to compute the current). It turns out that if one wants to
compute local quantities like the current or other local two-point functions
(see Fig.\ref{fXIII}), the grand canonical ensemble gives correct results. We have
to keep in mind that the finite-size effects are of order $L^{-1}$
(see Eq.(\ref{eqIII.XI})) as opposed to $L^{-2}$ for the canonical ensemble (see
Eq.(\ref{eqV.XXVI})). If, however, one is interested in the values of the correlation
functions at fixed  macroscopic distances $y$ (see the definition in
Eq.(\ref{eqIV.III})) using the grand canonical ensemble one gets wrong
results. In order to illustrate this point, in Fig.\ref{fXc} we compare the
values of the correlation function $c_{+,-}$ obtained using the grand canonical
ensemble and the canonical ensemble (Monte-Carlo simulations) in the mixed
phase. We notice that around $y=0$ or 1, the two methods give the same
results but that the results differ drastically otherwise.

Can the grand canonical still be used in order to get information about
the mixed phase? We believe that the answer is yes and we have explored
this possibility. The idea is inspired by the way the problem is solved for
the usual Bose-Einstein condensation (the procedure is shortly repeated
in Appendix C). There, one introduces in the Hamiltonian a supplementary
term, which breaks the $U(1)$ symmetry of the problem, proportional with a
constant $\nu$ and the square of the volume. After a Bogoliubov
transformation, the net effect is a modification of the partition
function given by the grand canonical ensemble by a factor which depends
on the volume, $\nu$ and the chemical potential. This changes the relation
between the density and the chemical potential. In order to compute a
physical quantity, one keeps $\nu$ fixed, then one takes the volume to infinity
and at the end, makes $\nu$ equal to zero. Since in our case we were not able to
find a way to break explicitely the $U(1)\times U(1)$ symmetry of the problem, we
have, in analogy with what is done for the Bose-Einstein gas, just
modified the the partition function (\ref{eqII.XXXII}) of the grand canonical ensemble
by a factor which takes into account that we have two chemical potentials
and that at the critical point the chemical potential is not zero:
\begin{eqnarray}
Z_L^{\mu_1,\mu_2}(\nu)=\exp(\frac{L\nu^2}{2\mu_c-\mu_1-\mu_2}) Z_L^{\mu_1,\mu_2}
\label{eqVIII.VI}
\end{eqnarray}
In Fig.\ref{fXXIX} we show the $L$ dependence of $\zeta$ defined by Eq.(\ref{eqVIII.II}) in the
mixed phase for various values of $\nu$. We have to keep in mind that the
range of values of $L$ is limited. For $\nu=0$, 0.05, 0.06 and 0.07 it looks
like the compressibility $\kappa$ (see Eq.(\ref{eqVIII.I}) diverges. A dramatic change
takes place for $\nu=0.08$, the compressibility becomes convergent. It is
plausible to assume that the compressibility converges for any
non-vanishing $\nu$ (one has to take larger lattices to observe the
phenomenon). We further investigate the effect of the modification of the
partition function. In Fig.\ref{fXXVIII} we show again the the correlation function
$c_{+,-}$ as a function of $y$ (compare with Fig.\ref{fXc}), but this time for various
values of $\nu$.

We notice that between $\nu=0.07$ and $\nu=0.08$ again a dramatic change takes
place: the correlation function (it is not the connected one!) drops to a
constant value. It is tempting to assume that this should be the
correlation in the fluid observed in the mixed phase (the
symmetry breaking term eliminates the condensate also in the case of the
Bose-Einstein gas). This would imply:
\begin{eqnarray}
c_{+,-}=R_+^{\rm fl} R_-^{\rm fl}=(R_+^{\rm fl})^2
\label{eqVIII.VII}
\end{eqnarray}
where $R_+^{\rm fl}$ is given by Eq.(\ref{eqV.XXXV}). Using Eq.(\ref{eqV.X}) this gives $c_{+,-}=0.031$.
This is the solid line shown in Fig.\ref{fXXVIII} which gives approximately the
correlation function observed for $\nu=0.08$.

   It is obvious that this investigation is only preliminary but we
believe that it gives the right approach to derive the properties of the
fluid in the mixed phase. It also shows how similar are the condensation
phenomenon in the Bose-Einstein gas and in our model. 

\section{Conclusions\label{seccon}}

We believe that we brought some more light in the understanding of phase 
transitions in one-dimensional stationary states. The model we considered
has two advantages: it has interesting physical properties and has 
behind it a mathematical structure given by a quadratic algebra. This 
mathematical structure is not only a useful computational tool but 
allows, due to its similarity to the mathematical formulation of 
equilibrium statistical physics, to define relevant quantities which 
characterise the nature of the phase transition. A first example is the 
definition of the grand canonical ensemble. Working with this ensemble 
has not only allowed to obtain exact results (like the large $q$ expansion
presented in Sec.\ref{secII_IV}) but also to do precise numerical calculations on 
large lattices. Moreover, it allowed to define the analogue of the pressure
which, amazingly, has the same properties as in equilibrium (see Fig.\ref{fXXX}).
Using the grand canonical ensemble we were able to identify the nature 
of the phase transition between the disordered and the mixed phase since 
this transition resembles the one seen in the Bose-Einstein condensation
(see the fugacity dependence of the density in Fig.\ref{fXXII} and the volume 
dependence of the compressibility in Fig.\ref{fXVI}). Most importantly, using the 
grand canonical ensemble, one can break explicitly the continuous 
symmetries of the problem (conservation of the number of charged 
particles) in analogy with the treatment of the same problem in the case of
Bose-Einstein condensation (see Eq.(\ref{eqVIII.VI}) and Figs.\ref{fXXIX} and \ref{fXXVIII}). 

 Before discussing the physics of the model, we would like to stress 
some other results presented in this paper.
 We have shown that the \FEF , introduced earlier in 
Ref.\cite{citU} 
to study first-order phase transitions, is a meaningful quantity also in the
present context. This was checked for two order parameters (see
Figs.\ref{fXVII}, \ref{fXIX} and \ref{fCI}) and the behavior of this function in the case of a
second-order phase transition was for the first time observed. The
\FEF\ can also be useful in order to put in evidence 
the breaking of translational invariance since in this case it does not
converge in the infinite volume limit. This observation is based only on the
data presented in Fig.\ref{fXVIII} and obviously one needs to do more work to 
clarify this point. 

 We think that we have also found the proper method to look at flip times
in order to distinguish between situations when one has charge segregation
with and without translational symmetry breaking. As described in Sec.\ref{secVI}, 
one has to look not only at the module of the order parameter defined by
Eq.(\ref{eqVI.I}) but also at its phase. 

Inspired by the numerical results presented in \ref{appMF}, in Sec.\ref{secIMF} we 
have shown in detail how to do mean-field calculations for this model. 
The present method can also be used for a 
larger class of models.
These mean-field equations are two coupled non-linear differential
equations (see \ref{appMF}). The fact that one is able to find analytic
solutions for the stationary states is very non-trivial. We will
return to this topic in Ref.\cite{citQ}
where we consider the case of different densities of positive and negative particles.
We will explain why the differential equations should be called the two-component
Burgers equations and give more analytic solutions of these equations.

 We found the physics of this simple model fascinating. There are two 
phase transitions. One separates the disordered and  mixed phases and is of
second-order. This phase transition is related to spatial condensation. We 
got estimates for 
several critical exponents related to the inverse correlation length, the
algebraic behavior of the two-point function at the phase transition and
finite-size scaling corrections (see Secs.\ref{secIMF} and \ref{secVII}). The mixed phase with its
simple expression for the current and its independence on the density and
other parameters as well as the \shocks\ well described by
mean-field are nice pieces of physics. The nature of the \shock\ is also 
interesting: one has two open systems glued together. One, the condensate
composed of charged particles is well understood from the corresponding
two-state problem. About the second, the fluid, we know little 
except for the fact that it shows no macroscopic structure. Actually, one major 
difference between Bose-Einstein condensation and spatial condensation 
is that here the condensate has a simple structure whereas the fluid 
phase is interesting since it is  different than
the disordered phase and hence worth studying. 

In another publication \cite{PVII} where we consider the open system with the same bulk rates
as in the present one we will show that a first-order 
phase transition occurs between one phase which is the condensate
and a second one which is the fluid. In this way one can get
directly more information about both phases.

Unfortunately we did not
pay much attention to the second phase transition, the one between the mixed
and pure phase, where translational invariance is spontaneously broken.

 This work has left us with quite a few open questions which we would 
like to address in some future publications:

(a) We have not touched the study of the model in the case of 
different densities. A second paper will be devoted to this. 

(b) The algebraic approach and the Monte-Carlo simulations were used only
for a limited numbers of points in the $q$--$\lambda$ plane. Most of our
conclusions including the transition separating the mixed phase form the
disordered phase are mean-field results which were therefore checked only 
in a limited region of the $q$--$\lambda$ plane. This does not mean that in other 
regions mean-field is valid.

(c) The dynamics of the model with the parameters chosen such that the 
stationary states correspond to the pure and mixed phases as well as 
at the phase transition points can lead to plenty of interesting 
results.     

(d) Last but not least, there is many information given in this paper
which deserve a more accurate research. We did our best but we feel that
this is not enough.

\ack

The authors would like to thank C.Godr{\`e}che for a substantial contribution
to this paper. They offered him to be a co-author but he graciously
declined. They would also like to thank O.Zaboronsky for asking a very
pertinent question concerning the solutions of the mean-field equations.
They are also grateful to F.C.Alcaraz, M.Evans, K.Mallick, and
A.Nersesyan for useful discussions. One of them (V.R.) would like to thank
SISSA, Trieste and the Einstein Center at the Weizmann Institute for their
warm hospitality.
This work was supported by the TMR Network Contract FMRX-CT96-0012 
of the European Commission.
\appendix
\section{The connection between the representations of the 
quadratic algebra and of the quantum algebra $U_qso(2,1)$. 
Recurrence relations.\label{appA}}

The expressions (\ref{eqII.VII})-(\ref{eqII.X}) of the representations of the quadratic algebra
(\ref{eqII.III}) suggest to any reader familiar with the representations of quantum algebras
a connection between these very different mathematical topics. This connection is for 
the purposes of this paper not only a mathematical curiosity but, as will be seen, 
one can write recurrence relations for physical relevant quantities which have an 
interpretation from the quantum algebras point of view. We will comment about these 
recurrence relations at the end of this Appendix.

We first remind the reader of the definition of the quantum algebra $U_so(2,1)$. 
This is the deformation given by a parameter $s$ of the well known algebra $so(2,1)$.
(The 
deformation parameter is usually denoted by $q$ but this symbol is already used in this
paper). The quantum algebra is defined by three generators $k$, $e$ and $f$ satisfying the 
relations \cite{citX}:
\begin{eqnarray}
[e,f]=\frac{k^2-k^{-2}}{s-s^{-1}}
\nonumber\\
k\,e\,k^{-1}=s^{-1} e
\nonumber\\
k\,f\,k^{-1}=s\, f
\label{A.1}
\end{eqnarray}
One representation of this algebra  has the following matrix elements:
\begin{eqnarray}
k=\left(\begin{array}{rrrrrrrrr}
             k_1&0&0&0&\cdots\\
	     0&k_2&0&0\\
	     0&0&k_3&0\\
	     0&0&0&k_4\\
	     \vdots&&&&\ddots
	\end{array}\right)
\qquad
e=\left(\begin{array}{rrrrrrrrr}
             0&u_1&0&0&\cdots\\
	     0&0&u_2&0\\
	     0&0&0&u_3\\
	     0&0&0&0&\ddots\\
	     \vdots&&&&\ddots
	\end{array}\right)
\nonumber\\
f=\left(\begin{array}{rrrrrrrrr}
	     0&0&0&0&\cdots\\
             v_1&0&0&0\\
	     0&v_2&0&0\\
	     0&0&v_3&0\\
	     \vdots&&&\ddots&\ddots
	\end{array}\right)
\label{A.2}
\end{eqnarray}
where 
\begin{eqnarray}
k_j=s^j\,,\qquad\qquad u_j\,v_j=[j]_s\,[j+1]_s
\end{eqnarray}
and
\begin{eqnarray}
[j]_s=\frac{s^j-s^{-j}}{s-s^{-1}}\;.
\end{eqnarray}
Notice that if
\begin{eqnarray}
s^2=r=\frac1{\lambda}
\end{eqnarray}
then $G_1$ and $G_2$ (see Eq.(\ref{eqII.VII})) can be expressed in terms of $k$, $e$ and $f$ as follows:
\begin{eqnarray}
G_1=\frac{k}{s}\left(e+\frac{k-k^{-1}}{s-s^{-1}}\right)
\,,\qquad
G_2=\frac{k}{s}\left(f+\frac{k-k^{-1}}{s-s^{-1}}\right)
\end{eqnarray}
In order to show how the connection between the representation of the quadratic algebra
and the quantum group can be used, we will consider a simpler example where $s=1$ (this 
is the familiar case of the $so(2,1)$ algebra). In this case, one has:
\begin{eqnarray}
G_1=S^z+S^-
\,,\qquad
G_2=S^z+S^+
\end{eqnarray}
with:
\begin{eqnarray}
S^z \,|k\!> &= k \,|k\!>
\nonumber\\
S^- \,|k\!> &= \sqrt{k(k-1)} \,|k-1\!>
\nonumber\\
S^+ \,|k\!> &= \sqrt{k(k+1)} \,|k+1\!>
\label{A8}
\end{eqnarray}
where $k\geq 1$.
We have to keep in mind (see Eq.(\ref{eqII.VIII})) that the matrix $G_0$ has a simple expression:
\begin{eqnarray}
G_0\,|k\!>=\delta_{k,1}\,|k\!>
\label{A9}
\end{eqnarray}
The calculation of the current for example, in the grand canonical ensemble, implies the 
calculation of the quantity (see Eqs. (\ref{eqII.XXXII})-(\ref{eqII.XXXIV})):
\begin{eqnarray}
C^N\,|1\!>
\end{eqnarray}
where
\begin{eqnarray}
C=G_0+(z_1+z_2)S^z+z_1S^-+z_2S^+
\end{eqnarray}
Writing $C^N \,|1\!>$ in the $|n\!>$ basis:
\begin{eqnarray}
C^N\,|1\!>=\sum_{n=1}^{N} t_n^{(N)}\,|n\!>
\end{eqnarray}
one can find a recurrence relation for the $t_n^{(N)}$ using (\ref{A8}) and (\ref{A9}). It is useful 
to denote:
\begin{eqnarray}
\sqrt{n}\; t_n^{(N)} = u_n^{(N)}
\end{eqnarray}
and we get the following recurrence relations for $2\leq n\leq N-1$:
\begin{eqnarray}
u_{1}^{(N+1)} = (1+z_1+z_2) \,u_{1}^{(N)} +z_1\, u_{2}^{(N)}
\nonumber\\
\frac1{n}\, u_{n}^{(N+1)}= (z_1+z_2)\, u_{n}^{(N)} +z_1\, u_{n+1}^{(N)} +z_2\, u_{n-1}^{(N)}
\nonumber\\
\frac1{N}\, u_{N}^{(N+1)}=  (z_1+z_2)\, u_{N}^{(N)}+ z_2\, u_{N-1}^{(N)}
\nonumber\\
\frac1{N+1}\, u_{N+1}^{(N+1)}= z_2\, u_{N}^{(N)}
\label{A14}
\end{eqnarray}
If in the recurrence relations (\ref{A14}) one drops the $1$ in the term $(1+z_1+z_2)$ one 
obtains recurrence relations defined only by the algebra $so(2,1)$. Similar recurrence 
relations can be found also in the deformed case ($s$ different from $1$). We just did not 
have the time to exploit the recurrence relations just mentioned, they might appear 
however in a further publication. 

Before closing this Appendix let us observe that for
\begin{eqnarray}
q=\frac1r=\lambda+1
\end{eqnarray}
$G_1$ and $G_2$ can be expressed through the generators of the $r$ deformed 
bosonic creation and annihilation operators:
\begin{eqnarray}
G_1=1+\sqrt{1-r}\,a
\,,\qquad
G_2=1+\sqrt{1-r}\,a^+
\label{A16} 
\end{eqnarray}
where
\begin{eqnarray}
a\,a^+ -r\,a^+\, a=1
\end{eqnarray}
and
\begin{eqnarray}
a\,|1\!>=<\!1|\,a^+=0
\end{eqnarray}
This allows again to write recurrence relations like (\ref{A14}) with a clear
mathematical meaning.

In defining the grand canonical ensemble (see Sec.~2) we had to make a choice between isomorphic algebras
(Eqs.(\ref{eqII.XXXIII})--(\ref{eqII.XXXIV})) corresponding to different normalizations
of the generators $G_0$, $G_1$ and $G_2$.
As we can see from Eqs.(\ref{A8}) and (\ref{A16}), the chosen normalisation corresponding to Eq.(\ref{eqII.XXX})
looks reasonable.

\section{\label{appMF} Mean-field on the lattice and in the continuum}

\fCX
\fCXI
\newcommand{\rhop}[1]{p_{#1}}
\newcommand{\rhom}[1]{m_{#1}}

Let $p_k$, $m_k$ and $v_k$ be the average concentrations on the site $k$ of positive.
negative particles and vacancies. Neglecting correlations, the expression 
of the currents of positive and negative particles on the link between the 
sites $k$ and $k+1$ are:
\begin{eqnarray}
J^+_{k,k+1}=q \, \rhop{k}\rhom{k+1} - \rhom{k}\rhop{k+1} + \lambda\, \rhop{k}(1-\rhom{k+1}-\rhop{k+1})
\nonumber\\
J^-_{k+1,k}=q \,\rhop{k}\rhom{k+1} - \rhom{k}\rhop{k+1} + \lambda \,\rhom{k+1}(1-\rhom{k}-\rhop{k})
\label{eqB.I}
\end{eqnarray}
The time evolution of the local densities can be derived from the master 
equation:
\begin{eqnarray}
\frac{\d}{\d t} \rhop{k}=J^+_{k-1,k}-J^+_{k,k+1} 
\nonumber\\
\frac{\d}{\d t} \rhom{k}=J^-_{k+1,k}-J^-_{k,k-1} 
\label{eqB.II}
\end{eqnarray}
For the stationary state one gets a system of nonlinear equations for 
the local concentrations which might give several solutions. In order to 
find them, we have proceeded as follows. We have used Eq.(\ref{eqB.II}) taking 
discrete values for the time and chose the parameters ($q,\,\lambda$ and $\rho$) 
corresponding to the pure and mixed phase. As an initial distribution we 
have put at random $\rho L$ positive and $\rho L$ negative particles on the lattice. 
Starting from one of these configuration the iteration converges either to the 
homogeneous solution ($p_k=m_k=\rho$) or to inhomogeneous solutions. The 
inhomogeneous solutions have all the same profiles which differ only in 
their position on the ring. Therefore, in mean-field, translational invariance is broken 
in both the pure and mixed phases even for finite $L$. In Figs. \ref{fCX} and \ref{fCXI}  the
inhomogeneous solutions are shown for the pure and mixed phase. In both 
figures one sees two domains. In one domain one has only charged particles  
with  non-uniform distributions (the condensate) and in the other domain one has uniformly 
distributed vacancies and charged particles (the fluid). 
For a given values of the parameters $q$, $\lambda$ and $\rho$, the
inhomogeneous solution appears only for a lattice size $L>L_{\rm min}$. For larger
lattice sizes, one finds inhomogeneous solutions different of those shown
if Figs.\ref{fCX} and \ref{fCXI}. Namely if $L>2 L_{\rm min}$ one finds solutions looking like two
pictures shown in Fig.\ref{fCXI} glued together (the same is true for the pure
phase). Otherwise speaking, for a lattice size $2 L$ one gets a solution
with two condensates, in each condensate one sees the same concentrations 
profiles as in the case of a single condensate for a lattice size $L$. The
reader can guess what happens for $L>3L_{\rm min}$. Based on this observations, we
now look at the mean-field equations in the continuum.

\newcommand{\pary}[1]{\frac{\partial {#1}}{\partial y}}
\renewcommand{\rhop}{\rho_+}
\renewcommand{\rhom}{\rho_-}
\newcommand{\rhon}{\rho_0}
Instead of the Eqs.(\ref{eqB.I}) for the currents, one gets:
\begin{eqnarray}
J^+=&(q-1)\rhop\rhom + \lambda \rhop\rhon 
\nonumber \\
&+\frac1{2L} \left( (q+1)(\rhop \pary{\rhom} -\rhom \pary{\rhop}) +
                  \lambda (\rhop \pary{\rhon} -\rhon \pary{\rhop}) \right)
\nonumber \\
J^-=&(q-1)\rhop\rhom + \lambda \rhom\rhon 
\nonumber \\
&+\frac1{2L} \left( (q+1)(\rhop \pary{\rhom} -\rhom \pary{\rhop}) +
                  \lambda (\rhon \pary{\rhom} -\rhom \pary{\rhon}) \right)
\label{eqB.III}
\end{eqnarray}
where $\rhop$, $\rhom$ and $\rhon$ are the densities of positive, negative
particles and vacancies, 
\begin{eqnarray}
\rhop(y)+\rhom(y)+\rhon(y)=1
\label{eqB.IV}
\end{eqnarray}
The Eqs.(\ref{eqB.II}) are replaced by:
\begin{eqnarray}
\frac{\partial \rhop}{\partial t}=- \pary{J^+}
,\qquad
\frac{\partial \rhom}{\partial t}=  \pary{J^-}
\label{eqB.V}
\end{eqnarray}
The solutions of the Eqs.(\ref{eqB.V}) have to be positive, fulfill Eq.(\ref{eqB.IV}) as
well as
\begin{eqnarray}
\int_0^1 \rhop(y) \,\d y =p
,\qquad
\int_0^1 \rhom(y) \,\d y =m
\label{eqB.VI}
\end{eqnarray}
Notice that solutions of the mean-field equations will depend on the
parameters $L,\,q,\,\lambda,\,p$ and $m$.

A more detailed study of the solutions of Eq.(\ref{eqB.V}) will be presented
elsewhere \cite{citQ}, in the present paper we are interested in the stationary
solutions which are obtained when $p=m$ 
and with only one condensate. These solutions are obtained in Sec.\ref{secIMF}
based on the observation that in the condensate one has only two species
which makes the problem solvable and that in the fluid the concentrations
are uniform. It turns out that in the presence of fluctuations only
the solutions with one condensate are relevant for the stationary states.

\section{\label{appBEC} Bose-Einstein condensation in equilibrium and spatial condensation in stationary states.}
\def\kv{\vec{k}}
\def\nt{\tilde{n}}

We will first review the phenomenon of condensation for the free Bose 
gas \cite{citY}. The partition function $Z$ of the system is:
\begin{eqnarray}
Z=\tr \, \exp -\beta\sum_{\kv}\left((\epsilon(\kv)-\mu)n_{\kv} - \sqrt{V} \nu (a_0+a_0^+)\right)
\label{eqC.I}
\end{eqnarray}
where $\beta$ is the inverse of the temperature, $\epsilon(\kv)$ is the energy, $\kv$ 
the momentum, $\mu$ the chemical potential, $V$ the volume, $\nu$ gives the 
intensity of the symmetry breaking term and $n_{\kv}$ is the occupation number:
\begin{eqnarray}
n_{\kv}=a_{\kv}^+a_{\kv} 
\label{eqC.II}
\end{eqnarray}
The $a_{\kv}$'s are Bosonic annihilation operators. We assume for simplicity that we 
consider the three-dimensional gas with a quadratic dispersion relation so that one has
Bose-Einstein condensation.
Since $\epsilon(0)=0$ we have a zero Bosonic mode. In the absence of the 
symmetry breaking term ($\nu=0$), the Hamiltonian is invariant under  
$U(1)$ transformations
\begin{eqnarray}
a_0 \rightarrow a_0 \,e^{\i \phi}
\label{eqC.III}
\end{eqnarray}
where $\phi$ is a phase. Through the non-unitary Bogoliubov transformation
\begin{eqnarray}
b_0=a_0+\sqrt{V} \frac{\nu}{\mu}
\label{eqC.IV}
\end{eqnarray}
the partition function becomes
\begin{eqnarray}
Z=e^{-\frac{\beta V \nu^2}{\mu}} \tr \,\exp -\beta\sum_{\kv}\left(\epsilon(\kv)-\mu)\nt_{\kv}\right) 
\label{eqC.V}
\end{eqnarray}
where $\nt_{\kv}$ are again number operators ($\nt_{\kv}=0,1,2,\dots$)
\begin{eqnarray}
\nt_{\kv}=b_{\kv}^+b_{\kv} 
\label{eqC.VI}
\end{eqnarray}
The net effect of the Bogoliubov transformation is the appearance of the 
chemical potential not only under the trace operation but also in the 
overall factor. This implies that $\rho$ defined by the expression
\begin{eqnarray}
\rho=\frac1{V\beta} \,\frac{\partial}{\partial \mu} \log Z
\label{eqC.VII}
\end{eqnarray}
is the average density of particles only for $\nu=0$. The main point is that we 
first do the calculation for $\nu$ different from zero, take the infinite 
volume limit and only at the end let $\nu$ vanish. In this way one gets 
the thermodynamical properties of the gas also under the critical 
temperature since the condensate (which is a state of zero momentum) 
disappears from the calculation as a consequence of the symmetry 
breaking term. This method can also be seen as a regularization procedure.
 
That the free Bose gas is a quantum system (the expression (\ref{eqC.V}) 
contains only occupation numbers which are $c$-numbers) can be seen once we 
compute measurable correlation functions like
\begin{eqnarray}
g(\vec{r}_1,\vec{r}_2)=<\!\psi(\vec{r}_1)\psi^{\dag}(\vec{r}_2)\!>
\label{eqC.VIII}
\end{eqnarray}
where
\begin{eqnarray}
\psi(\vec{r})=\frac1{\sqrt{V}}\sum_{\kv}e^{\i \kv \vec{r}} a_{\kv}
\label{eqC.IX}
\end{eqnarray}
since the creation and annihilation operators verify the Heisenberg algebra.
That much on the Bose-Einstein gas.
 
In the case of the spatial condensation which appears in the mixed 
phase, the condensate is separated in space from the fluid and not in 
momentum space. 
The simplest model of spatial condensation that we know of is the model 
with a ``defect'' of Ref. \cite{citW}. In this model $M$ particles hop in 
a totally asymmetric way  with a rate 1 on a ring with $L$ sites and one 
particle (the ``defect'') hops with a rate $p$ (with $p<1$). It was shown that the
partition function for the grand canonical ensemble is
\begin{eqnarray}
Z=\sum_{n_0,n_1,\dots} \delta(\sum_{k=0} n_k - L) \frac1{p^{n_0}} \exp \left(\mu {\sum_{k=0} n_k}\right)
\label{eqC.X}
\end{eqnarray}
where $n_0,n_1,n_2,\dots$ are the number of vacancies in front of the 
defect, of particle 1, particle 2 etc.
Obviously,
\begin{eqnarray}
L-(M+1)=
\frac{\partial}{\partial \mu} \log Z
\label{eqC.XI}
\end{eqnarray}
At low density of particles (large density of vacancies), the particles 
produce a jam behind the defect and the string of vacancies (of length $n_0$) in front of the defect can be seen as a condensate. 
The expression (\ref{eqC.X}) looks similar to (\ref{eqC.I}) with at least two major 
differences: the occupations numbers $n_k$ are constrained by the length $L$
of the chain (we have condensation in real space and not in momentum 
space!) and there are no measurable correlation functions given by 
creation and annihilation operators in this problem.

In the model discussed in present paper, the analogy between the expression of
the partition functions (\ref{eqII.XXXII}) for the stationary state and (\ref{eqC.I}) for the Bose
gas is completely obscure. Some similarities do exist. The algebra (\ref{eqII.II}) 
plays a role similar to the Heisenberg algebras hidden in the occupation
numbers appearing in the Bose gas and the transformation (\ref{eqII.IV}) representing
the conservation of the numbers of positive and negative particles (we 
have a $U(1)\times U(1)$ symmetry in our case) corresponds to the 
transformation (\ref{eqC.III}). We were not able to bring more understanding 
in the symmetry breaking mechanism using the algebraic approach
although we wish we had, since we were looking for an equivalent of the 
Bogoliubov transformation in oder to eliminate the condensate. Our 
approach using Eq.(\ref{eqVIII.VI}) can be seen at this stage as an educated guess 
to do the regularisation.

%

%
\end{document}